**Title:** Associational and plausible causal effects of COVID-19 public health policies on economic and mental distress.

Reka Sundaram-Stukel[1] and Richard J Davidson[2]

## Summary
### Background
The COVID-19 pandemic has increased mental distress globally. The proportion of people reporting anxiety is 26%, and depression is 34% points. Disentangling associational and causal contributions of behavior, COVID-19 cases, and economic distress on mental distress will dictate different mitigation strategies to reduce long-term pandemic-related mental distress.

**Methods** We use the Household Pulse Survey (HPS) April 2020 – February 2021 data to examine mental distress among U.S. citizens attributable to COVID-19. We combined HPS survey data with publicly available state-level weekly: COVID-19 case and death data from the Centers for Disease Control, public policies, and Apple and Google mobility data. Finally, we constructed economic and mental distress measures to estimate structural models with lag dependent variables to tease out public health policies' associational and causal path coefficients on economic and mental distress.

### Findings
From April 2020 to February 2021, we found that anxiety and depression had steadily climbed in the U.S. By design, mobility restrictions primarily affected public health policies where businesses and restaurants absorbed the biggest hit. Period t-1 COVID-19 cases increased job-loss by 4.1% and economic distress by 6.3% points in the same period. Job-loss and housing insecurity in t-1 increased period t mental distress by 29.1% and 32.7%, respectively. However, t-1 food insecurity decreased mental distress by 4.9% in time t. The pandemic-related potential causal path coefficient of period t-1 economic distress on period t depression is 57.8%, and anxiety is 55.9%. Thus, we show that period t-1 COVID-19 case information, behavior, and economic distress may be causally associated with pandemic-related period t mental distress.

### Interpretation
Exploring the association and potential causal relationships between economic distress at t-1 and mental distress at t may have several interpretations. First, long-term pandemic-related involuntary job loss could have persistent wage scarring effects for many workers post-pandemic. Second, labor productivity may decline because of the lingering effects of mental distress. Third, as the economy recovers, investments in job-retraining programs with skill appreciation complemented with resilience training for displaced workers would enhance welfare. Finally, expanded mental health teleservices for workers adjusting to post-pandemic work-life may increase emotional well-being.

**Funding** No funding was used to support this research.

### Acknowledgements
We thank Dr. Adams for many discussions, valuable feedback, and detailed comments. Ms. Stukel for editing. All errors are our own.

### Author Information
[1]Dr. Reka Sundaram-Stukel (corresponding author)
Research Fellow, Department of Economics
7415 Social Sciences, University of Wisconsin–Madison, Madison, Wisconsin 53706
rsundara@wisc.edu

[2]Dr. Richard J Davidson
William James Professor of Psychology and Psychiatry
318 Psychology, University of Wisconsin–Madison, Madison, Wisconsin 53706
rjdavids@wisc.edu

## Research in context

### Evidence before this study

Our search terms in APA, PubMed, medRxiv, arXiv, NBER, and Econlit included “anxiety,” “fear,” “depression,” “mental health,” “COVID-19”, “COVID-19”, “economic fallout,” “stimulus,” “unemployment insurance,” “employment,” “eviction,” and “economic distress.” We focused on all studies from April 2020 to May 2021 that met the search criteria and were COVID-19 relevant. We found five major empirical studies that directly addressed either significant mental health or economic consequences of COVID-19. We excluded the studies that relied on simulated data or studies that used theoretical models unsupported by data. Nevertheless, there is a consensus that mental distress restricts consumption behavior, that mental distress is rising among the US population, those in high contact sectors disproportionately feel economic fallout, and that recovery from the COVID-19 pandemic is likely to follow a “K” shaped path where some economic sectors recover fast, and others fall behind. The mental health side of the story acknowledges that there are widespread increases in anxiety, fear, worry, and depression. Still, no large-scale studies explicitly link these factors to economic distress.

### Added value of this study

The unique large-scale shock of COVID-19 that precipitated stringent public health containment measures caused an economic and mental health crisis. We unpack the causal implications of case information, behavior, and public policy on economic and mental distress.

### Implications of all the available evidence

Compared to pre-pandemic reported prevalence rates for anxiety and depression, the national mental distress at the onset and during COVID-19 is alarming. Four critical mental health questions, highly correlated with generalized anxiety and depressive disorder, afflict 34% over half the days of the week, and an additional 12% reported anxiety or depressive symptoms for some days of the week. Thus, public health and economic policies should focus on minimizing both long-term wage and psychological scarring.

## Introduction

Little is written about the causal effects of COVID-19, the public health response, and the consequential economic distress on mental health. Because of the size of the COVID-19 pandemic, the road to recovery is going to be long and highly unequal. According to the IMF: World Economic Outlook, the current economic challenges in order of priority are equitable access to vaccines, safety nets for workers in affected countries and sectors either through retention or reallocation, a regulatory eye on inflation with strong fiscal and monetary instruments ready to engage, and restoring the GDP growth to pre-COVID-19 levels[1]. None of these global strategies include the co-occurring mental health crisis or how it may affect economic and psychological recovery in a post-pandemic world. The mental health fallout from the pandemic is significant and severe. A global mental health crisis existed before COVID-19, and post-pandemic, it may worsen for some countries without any aid[2-3]. Preliminary signs show that mental health crisis is likely more severe in low and middle-income countries (LMIC)[4-5]. It seems fitting that advanced economies should first explore the real consequences of mental distress on recovery.

While pre-COVID-19 studies have shown that mental distress results in significant losses due to absenteeism (days taken off) and presenteeism (at work but unproductive), we do not have data to measure the effects of COVID-19 related labor disruptions on worker productivity[6-9]. Brand J. (2015) offers a review where many studies show that the long-term effects of involuntary job-loss have irreparable economic and psychological consequences[10]. In addition, many studies have highlighted the mental health fallout from COVID-19[11-18]. Giuntella, O. et al. (2021), using data from U.S. college students, report significant disruption in routine behaviors such as physical activity and sleep that contribute to mental distress[11]. Witteveen, D., and Velthorst, E. (2020), focusing on European countries, find that the early economic consequences of COVID-19 lockdown are highly correlated with anxiety, depression, and loneliness[17]. Cobb R.J. et al. (2021), using America Trends Panel COVID-19 study, argued that these mental health consequences might be worse among the Black population[18]. Donnelly R, Farina MP. (2021) found that in the first 12 survey weeks of the household pulse survey (HPS), depression and anxiety prevalence varied significantly by state and by household income shocks. If a state had more comprehensive social policies like Medicaid expansion, unemployment insurance, and COVID-19 related utility abatement, the correlation between mental health and income shocks was weaker[14].

The National Health Interview Survey (NHIS) 2019 data reported that only 6.1% of Americans reported being moderately or severely anxious or depressed[18-19]. However, during August 2020 – February 2021, 36.4 to 41.5% of Americans had moderate to severe anxiety or depression[20-22]. Thus, even if the economy returns to normal employment levels and economic activity, the mental health fallout will likely linger and slow down economic recovery globally.

Chernozhukov, V. et al. (2021) provide a comprehensive causal exploration into public health policies, behavior and found that masking reduced COVID-19 cases between 19 to 47%. Their counterfactual analysis showed if businesses had remained open, it would have resulted in nearly 17 to 78% more cases; not having stay-at-home orders would have increased cases by 6 to 63% by June 2020. They also found people voluntarily reduced visits to workplaces, retail and grocery stores, and limited public transit use. More importantly, Chernozhukov V. et al. (2021) provided a framework to evaluate the causal effects of public policies[23-24].

While their study did not consider mental distress, we use a similar approach to evaluate the causal effects of case counts, public health policies, and economic distress on mental health[24-26]. We start by estimating a reduced-form ordered probit model. Next, we examine the correlates of mild to severe anxiety and depression among 1.8 million Americans responding to 24 HPS survey weeks. We then use a causal framework to tease out the causal path coefficients of mental distress accruing from the public health response and the economic fallout[24-26]. Finally, we estimate the models in parts using both COVID-19 cases and deaths as sources of input. We also use Google and Apple mobility data as inputs for the public health response.

We find that case information and economic distress are the primary causes of U.S. pandemic-related mental distress. Between the two, economic distress has a stronger causal impact between 56.9% to 57.8% of mental distress, and lagged case information is between 5.6 and 10.1% of mental distress. This is not surprising because the pandemic has unleashed uncertainty in our lives. We used 24 survey weeks that covered a period of April 2020 to February 2021. Of the three economic distress variables we used, job-loss had a 29.1% impact on mental distress, food insecurity impacted mental distress by -4.9% and housing insecurity by 32.7%. Mobility or voluntary mobility restrictions also are correlated with mental distress by 21.6% increase in mental distress. The causal relationship between economic and mental distress is important because it will dampen economic recovery if it persists. Mental distress has been delegated as a separate phenomenon cooccurring during the pandemic policy, and priority interventions have focused on containment measures, vaccine distributions, and economic safety nets. While none of these should be omitted, additional funding resources and explicit policies to promote worker health and mental wellbeing are needed—both in the US and globally.

## Methods

In partnership with five other agencies in early April, the Census Bureau developed the household pulse survey (HPS). The HPS was designed to capture household experiences during the pandemic[21-22]. The sample for the HPS is drawn from the Census Bureau Master Address File and the Census Bureau Contact Frame. We utilize data spanning April 23rd, 2020, to March 1st, 2021, for a total of 24 weeks for this study. The survey was done in three phases: phase 1 of HPS spanned April 23rd to July 21st, 2020 and was conducted weekly; phase 2 spanned August 19th, to October 26th, 2020, and was conducted bi-weekly but still retained the name weekly pulse to be consistent with phase 1; and phase 3 spanned October 28th and ended March 29th, 2021[22]. The majority of HPS respondents (80%) were asked to complete the survey only once, 15% were asked to complete it twice, and 5% were asked to complete it three times[22]. Because so few participants completed more than one weekly survey, we treat the weeks as repeated cross-sections. We retained 1.8 million observations across the 24 survey weeks. We deleted observations that had missing values for any of our key covariates of interest. This allowed for a cleaner estimation of models.

In this cross-sectional data, we merged weekly HPS data with the weekly state-level variables: case count from the CDC[27], mobility data from Google and Apple[28-29], and health policy changes[30]. In addition, for baseline measures of mental distress (MD), we use NHIS because the questions in HPS on mental health and food security are borrowed from there[18-19].

From the HPS survey, we use four mental health questions (anxiety, worry, interest, and feeling down) along with race, ethnicity, education, and pandemic-related job-loss. Job-loss takes on the value 1 if a person reported pandemic related loss of workQW. We also construct two variables that measure economic distress (ED). Housing insecurity takes on the value 1 if people are not confident they will meet their rent or mortgage. Food insecurity takes on the value 1 if respondents reported that they could not purchase enough food or were not confident they could buy what they needed. Combined, job-loss, housing insecurity, and food insecurity make up our key indicators of economic distress. The four mental health questions from the 2-item generalized anxiety (GAD2) and 2-item depression disorder scale (PHQ2) combined the 4 items from a mental distress (MD4) scale with a Cronbach alpha of 0.91. The scale takes on the value 1 (no symptomatic days) to 4 (symptomatic every day). We also create a binary mental distress variable which takes on the value 1 if people report experiencing symptoms for half or more days of the week (see annexed supplement for a summary of covariates eTables 1- 2).

NHIS does not provide data at the state-level because of the small sample size, so it is not directly comparable to HPS. In our estimations, we use state-level data and the aggregate number of cases and deaths by weeks as they appear in the HPS survey[22]. Additionally, aggregating by state and HPS week, we create data on the percent change in mobility along six dimensions using Apple and Google mobility data and public policy indicator variables using Raifman et al. (2021)[30]. The mobility variables measure relative changes in visits (behavior) at time *t-1* (previous week) to recreational spaces, parks, transit stations, grocery or retail stores, pharmacies, and workplaces; the apple mobility data measures the relative change in driving patterns. The policy variables considered are safer-at-home, closures of essential businesses (daycare, business, restaurant), masking, and job-search requirements.

First, we uncover the correlates of mental distress by estimating an ordered probit model with MD4, GAD2, and PHQ2 using the full sample of HPS correcting for state and week fixed effects. In our analysis, this is the only model we estimate at an individual level. The rest of our models are estimated with aggregated data by state and HPS survey week. Next, we uncover the causal path coefficients of public health and economic distress on mental distress using a directed acyclic graph (DAG), structural equations, and an identification strategy similar to Chernozhukov V. et al. (2020) and Pearl (2000)[24-26]. To do so, we estimate several models in parts. The pandemic has moved at a swift pace. Governments implemented public health policies in sequential order, and they brought economic activity to a grinding halt. With this came elevated mental distress in the form of anxiety, loss of interest, worry, and depression. We conjecture that this rise in mental distress is potentially caused by public health policies and consequential economic distress. Figure 1 shows a more detailed DAG where period t-2 case information sets observed public health policy at t-1. The causal pathway is more complex, so we delineate the estimation strategy in parts with the following assumptions.

1. Mental distress at time t is determined last after case information, public health policy, mobility behavior, and economic distress are determined at time t-1.
2. Information at time t-2 determines case information and public health policy at time t-1.
3. Public health policy at t-1 determines behavior at t-1 (red dotted lines 1, 2, and 3 in Figure 1).
4. Economic distress is observed at t-1 after behavior and case information at t-1 (line 4).
5. Finally, mental health outcome is determined at t (line 5).

Figure 1 shows the sequence of decisions for causal estimations ($I_{t-2} \rightarrow PH_{t-1}, I_{t-1}$) in the post-COVID-19 world. Where case information $I_{t-2}$ at time t-2 affects public health policies at t-1 ($PH_{t-1}$) and case information at time t-1 ($I_{t-1}$) and public health policies affect behavior at t-1 $PH_{t-1} \rightarrow B_{t-1}$. Note we block a feedback loop where, $I_{t-2} \rightarrow B_{t-1}$, and $ED_{t-1}$. Case information at period t-1 affects economic distress in t-1 and together affects mental distress in period t. It is also possible for t-2 case information to affect mental health via a backdoor relationship (see eFigure 1h). Let us unpack this a little because case information is highly correlated with mental distress; there may be a causal backdoor path (highlighted in red). This would make sense because mental distress can have lingering effects from previous periods and worsen over time with case information.

These assumptions allow us to estimate the five distinct paths on the DAG carefully (Figure 1). The first model estimates the effect of information and public health policy on behavior (travel, restaurant, …) ($I_{t-2} \rightarrow PH_{t-1} \longrightarrow B_{t-1}, I_{t-2} \rightarrow I_{t-1}$). The second model estimates the effect of behavior and information on economic distress $I_{t-1} \rightarrow ED_{t-1} \leftarrow B_{t-1}$. And the third model estimates the effect of information, public health policy, and economic distress on mental health($I_{t-1}, \rightarrow MD_t \leftarrow I_{t-2}, ED_{t-1} \rightarrow MD_t$). This gives rise to a system of equations that are estimated using a structural model (see annexed supplement for equations). We do not incorporate higher-order time effects in our model because their effects on mental distress are likely to be pretty small. We also do not include confounders like political unrest that created fluctuations in mental health because while the political climate affected everyone, it is captured by the behavior covariates entirely. Once estimated, we can determine the causal path coefficients and have clarity on two things. Did the public health policies have a higher direct effect on mental distress, or did economic distress?

## Results

In Table 1, the results for an ordered probit estimation, a reduced form analysis, for all household pulse survey respondents show that log cases from the previous survey-week have a 2.4% effect on mental distress. However, the resulting economic distress has much larger effects. Pandemic-related mental distress likelihood increased by 34.5% for respondents who experienced job-loss, 36.4% of respondents were housing insecure, and 62.6% if they were food insecure. Women are 24.5% more likely to experience mental distress than men. Compared to those with high school and some college experience, those with a Bachelor's degree or more are 5.4% more likely to experience mental distress. Within race classifications, as compared to Whites, all races, except American Indian and Native Alaskans, are 6% more likely to experience mental distress, are less likely to experience mental distress. The cutoff points show that the boundary between healthy and some bad days is z=0.398; the boundary between half bad days and almost every day in mental distress has a z-score of z=1.36, which corresponds to 20% of the sample. The correlation matrix of economic distress and mental distress is available in the annexed supplement (eTable1). The causal relationship between mental distress and economic distress is complex because time t-1 COVID-19 cases, behavior, and public policies also have causal relationships with mental distress. As discussed above, the public policies, behavior, and case information are available at the state-level but not at an individual level. Tables 2 through 4 present the results, aggregated at the state-level, for the potential causes of mental distress. In all estimations, we included state and week dummies.

First, according to NHIS, in 2019, between 6 to 9% of Americans reported being anxious or depressed[18-19]. While these estimates are not directly comparable, they give us a general idea that mental distress stemming from COVID-19 is a serious public health concern. Second, Figure 2A shows how mental distress has varied over 24 survey weeks of the pandemic from April 23rd, 2020, to February 2021. There is a steady climb in the proportion experiencing mental distress for all states irrespective of high (red line) or low (blue line) cases. Figure 2B shows a comparison of 1st quarter 2019 against 4th quarter 2019 of NHIS data, and we can see that the proportions that were experiencing mental distress remain more or less the same. Figure 2C presents 2nd quarter 2020 against 1st quarter 2021 and shows that the proportion in mild mental distress (in 7 days) is significantly higher in the 1st quarter of 2021 irrespective of case levels. Figure 2D repeats this graphic for those experiencing moderate to severe mental distress in 7 days, and we find the same results. Thus, Figures 2A-D point to a sustained state of pandemic-related mental distress irrespective of covid-19 caseload.

In Table 2, we present the aggregate effects that period t-1 COVID-19 case information has on public health policy and the effects that public policy has on behavior in the same period. Overall, we find COVID-19 cases at t-2 have a large positive and significant relationship with public health policies at t-1—public mask mandates, unemployment insurance, and restaurant closures with the biggest positive coefficients. This is as expected as COVID-19 cases increase containment policies become mandatorily or voluntarily enforced. The overall effect of public health policies at t-1 on behavior at t-1 is negative, meaning mobility decreases by 26%. Masking is significantly and negatively correlated with the grocery and pharmacy mobility index (Table 2: $\sum \beta_{PH,B}$).

In Table 3, we present the results of estimating the effects of behavior and case information on economic distress. First, COVID-19 cases at t-2 increase COVID-19 cases by 52%, and behavior at t-1 decreases COVID-19 cases at t-1 by 10%. This is consistent with the weekly volatile nature of COVID-19 cases and resulting public policies and behavior. Next, COVID-19 cases at t-1 increase pandemic-related job-loss at t-1 by 4.1%. Thus, the total direct effect of COVID-19 cases at t-1 on economic distress at t-1 increases by 6.3% (Table 3, $\beta_5$). Even though this is an obvious result, it is good to see it because that means the model is not producing spurious results. Finally, behavior at t-1 decreases COVID-19 cases by 10% (Table 5, $\beta_4$) and increases economic distress by 8% (Table 5, $\beta_6$). Together, these data show that COVID-19 cases and public health policies at t-1 had a significant negative economic impact for the same period.

In Table 4, we present results of estimating the effect of COVID-19 cases at t-1, economic distress at t-1, and behavior at t-1 on mental distress at t. The annexed supplement presents the correlation between mental distress and covariates used (eTable 7-9). COVID-19 cases and economic distress at t-1 causally increase mental distress at t by 5.6% and 56.9% respectively (Table 4 MD Continuous Long 2). Separating the overall causal effect of economic distress at t-1 on mental distress at t, we can see that it lies between 55.9 to 57.8% for anxiety and depression, respectively. Dissecting individual components of economic distress at t-1, we can see that job-loss increases mental distress at t by 29.1%, and housing insecurity at t-1 increases mental distress at t by 32.7%. These results differ across mental afflictions—job-loss increases depression more by 3.9% than anxiety, and housing insecurity increases anxiety more by 2.6% than depression like symptoms.

Finally, we report the causal path coefficients for case information, behavior, and economic distress on mental health. Separating each of these out is important because we want to understand if the mental health effects will linger after the economy returns to normal. In the annexed supplement, we provide the system of equations that calculated the path coefficients for the DAG in Figure 1. Table 5 presents the path coefficients; as expected, COVID-19 case information decreases mobility by 26.0%, and t COVID-19 cases increased t-1 COVID-19 cases by 52.2%. Because public health policies closed down essential businesses, we can see that behavior at t-1 increased economic distress by 8.0%, and the direct effect of COVID-19 cases at t-1 on economic distress at t-1 is 6.3%. The path coefficients for mental distress were as follows: Covid-19 case information at t-1 increases mental distress at t by 11.9%, time t-2 case information has a direct effect of increasing mental distress by 6.3% (which tells us that mental distress does dissipate period to period) for a cumulative effect of COVID-19 cases on mental distress of 5.6% points. Behavior at t-1 increases mental distress by 21.6% and economic distress increases mental distress by 56.9%. As we can see, pandemic-related economic distress has the highest impact on mental distress.

When we disaggregate mental distress into anxiety and depression, we can see the path coefficients of economic distress on depression are 1.9% higher than anxiety. Similarly, case information path coefficients to depression are 8.2% lower than in anxiety. Thus, we can conclude that COVID-19 case information increases anxiety more than depression. This finding has implications that we take up in the discussion section. We used death as input for sensitivity analysis and found similar results; overall, all coefficients had a lower magnitude but similar signs. The coefficients of significant difference were that COVID-19 related death information at t-1 had decreasing effects on mental distress at t, and behavior restrictions at t-1 significantly decreased deaths by 21%. Analysis for the tables with death cases is in the annexed supplement in eTable3-eTable5. Additionally, a summary of variables(eTable1), correlation analysis for mental health NHIS and HPS (eTable2), and correlational analysis(eTable6-8) are in the annexed supplemental.

## Discussion

In this paper, we sought to unearth the correlation between economic distress and mental health. We estimated a reduced form ordered probit model and found that economic distress (measured by food security, housing security, and pandemic related job-loss) is highly correlated with psychological distress. Graphing the proportion in severe mental distress (more than half days bad), we saw an increase across the 24 survey weeks, which took us to February 2021. Our question was can we separate the contributions of public health policy and economic distress to mental distress, and was there more than an associational inference—that is, was it possible to make causal inferences between economic distress and mental distress?

To do that, we drew DAGs to guide our causal assumptions, and we imposed a sequential order, such that behavior, case information, public health response, and economic distress were determined before mental distress. Next, we began piecemeal estimations to recover causal path coefficients. We found that the direct effects of case information are highest on economic distress, then on public health. Behavior, as measured by Google and Apple mobility measures, is fully blocked by the public health containment measures by assumption. As we would expect, it has a strong negative impact on public health. These negative effects are driven by business, restaurant closures, and the presence of job search requirements. COVID-19 case information at t-1 is a strong driver of economic distress at t-1, and behavior restrictions at t-1 have an increasing effect on economic distress at t-1. Cumulatively COVID-19 case information at t-1 increases mental distress at t by 5.6% and economic distress at t-1 by 8%. These results are sensitive to model specification because of potential backdoor relationships with mental distress. For example, in a model where we add behavior, we find that mobility restrictions at t-1 increase mental distress at t by 21.6% (though this has a backdoor path so closer to 10%). Job-loss has the biggest effect on mental distress, between 27.3% for anxiety and 31.0%, for depression. Depending on the model specification we use, the causal effect of economic distress on mental distress is bounded between 55.9 to 57.8%.

Insights from previous studies show a high level of mental distress over short-term job-loss is concerning at three levels. First, because of its limiting effect on career progression[9-10]. Second, evidence shows wage hits persist well beyond the economic downturn. For example, Couch & Placzek (2010) reported wage scarring where immediate job-loss results in 33% wage cuts and 15% wage cut six years post initial job-loss[8]. Third, if persistent like depression, mental distress would impair a displaced workers' ability to capitalize on job retraining programs or impair a worker's ability to be productive once the economy returns to normal[6,7]. This would be particularly true if workers and firms do not stay matched. Our results fit in line with both the pandemic related COVID-19 studies on mental distress and economic distress. Finally, disaggregating the effects of mental distress into anxiety and depression separately, we see that those who report depression like symptoms are more deeply affected by UI ending, job-loss, food, and housing insecurity. Thus, this group could benefit from resilience training interventions.

Evidence from the United Kingdom shows a strong correlation between pandemic related economic distress and mental distress[11]. The findings from other mental health studies all report high levels of COVID-19 related anxiety and depression among Americans[13-17]. Our results show that the causal interpretation of public health containment on economic distress is negative, and it is positive on mental distress. However, the causal effects of economic distress on mental distress far outweigh the effects of public health containment measures on mental distress. It is this finding that requires policy attention. Brand J. 2015 gives a review of the long-run effects of job-loss and unemployment. She provides evidence that involuntary unemployment leaves a scarring effect both on wage and career progression and on psychological health[9-10].

In the US context, the American Rescue Plan Act (which provides additional relief to address the economic impact of COVID-19) passed the house and senate majority and became law in March 2021[32]. Other advanced economies have pursued similar stabilization fiscal policies. Still, according to the HPS data, post-survey weeks reported in this analysis, mental health continues to afflict a significant proportion of Americans. With the mutant variants threatening to destabilize economic activity again, it is a more urgent public health concern to pay attention to the long-run effects of mental distress. These results are generalizable to all countries affected by COVID-19, as current evidence shows global mental health concerns are widespread [2-3]. The causal inference of economic distress being the dominant driver of mental distress should be a wake-up call for all public health policymakers.

## Table 1. The correlates of mental distress using ordered probit model and full HPS sample*

| Dependent Variable→ | MD 4[1] | GAD 2[2] | PHQ 2[3] |
|---|---|---|---|
| Log cases$_{t-1}$ | 0.024*** | 0.020*** | 0.022*** |
| | (0.002) | (0.002) | (0.002) |
| Female | 0.256*** | 0.163*** | 0.305*** |
| | (0.007) | (0.007) | (0.006) |
| Age | -0.010*** | -0.009*** | -0.011*** |
| | (0.000) | (0.000) | (0.000) |
| **Education[4]** | | | |
| Less than High School | 0.003 | 0.026 | -0.016 |
| | (0.025) | (0.025) | (0.024) |
| Bachelors + | -0.054*** | -0.116*** | 0.028*** |
| | (0.006) | (0.006) | (0.006) |
| **Race[5]** | | | |
| Hispanic | -0.109*** | -0.094*** | -0.115*** |
| | (0.011) | (0.011) | (0.011) |
| Black | -0.160*** | -0.124*** | -0.186*** |
| | (0.008) | (0.009) | (0.010) |
| Asian | -0.140*** | -0.098*** | -0.200*** |
| | (0.017) | (0.018) | (0.016) |
| AIAN | 0.064*** | 0.062*** | 0.049*** |
| | (0.013) | (0.015) | (0.013) |
| **Economic Distress** | | | |
| Job Loss$_{t-1}$[6] | 0.345*** | 0.312*** | 0.349*** |
| | (0.007) | (0.007) | (0.007) |
| Housing Insecure$_{t-1}$[7] | 0.364*** | 0.335*** | 0.355*** |
| | (0.011) | (0.012) | (0.011) |
| Food Insecure$_{t-1}$[8] | 0.626*** | 0.608*** | 0.588*** |
| | (0.012) | (0.013) | (0.012) |
| Cut off point 1 | -0.285*** | -0.262*** | -0.368*** |
| | (0.023) | (0.029) | (0.022) |
| Cut off point 2 | 0.762*** | 0.750*** | 0.628*** |
| | (0.019) | (0.023) | (0.018) |
| Cut off point 3 | 1.472*** | 1.371*** | 1.222*** |
| | (0.019) | (0.023) | (0.018) |
| Observations | 1,809,043 | 1,809,043 | 1,809,043 |

Source: Own estimations using Household Pulse Survey
Robust standard errors in parentheses
Statistical significance levels*** p<0.01, ** p<0.05, * p<0.1
*Interpretation: per unit change in covariate the coefficient ($\beta \times 100$) is the percent change.
[1]Mental Distress 4-item scale with values 1-4 where 1= no days of mental distress, 2=few days of mental distress, 3=over half days of mental distress, and 4 = every day in mental distress.
[2]GAD2 2-item generalized anxiety scale with values 1-4 where 1= no days of anxiety, 2=few days of anxiety, 3=over half days of anxiety, and 4 = every day in anxiety.
[3]PHQ2 2-item depression scale with values 1-4 where 1= no days of depression, 2=few days of depression, 3=over half days of depression, and 4 = every day in depression.
Reference categories: [4]High school + education; [5]White
[6]Have not worked in the past 7 days
[7]Not confident can pay mortgage or loan
[8]Not enough to eat or lack of choice in food

## Table 2: Aggregate effects of information (log cases)$_{t-2}$, behavior$_{t-1}$[1] on public policies$_{t-1}$ by state*.

**2a.** **Y = Log Cases $I_{t-2}$→$I_{t-1}$, Public Health Policy $I_{t-2}$→$PH_{t-1}$**

| Dependent Variable → | Log Cases $I_{t-1}$ | SAHO[2] | Daycare | Business | Religious | Mask | Restaurant | UI Ends[3] | Tele Rx |
|---|---|---|---|---|---|---|---|---|---|
| X=Log Cases $I_{t-2}$ | 0.522*** | | | | | | | | |
| | (0.044) | | | | | | | | |
| | | -0.159 | 0.034 | 0.236 | 0.146** | 1.237*** | 0.763*** | -0.000*** | 0.574** |
| | | (0.140) | (0.026) | (0.278) | (0.073) | (0.201) | (0.274) | (0.000) | (0.251) |
| Constant | Yes | Yes | Yes | Yes | Yes | Yes | Yes | Yes | Yes |
| State Dummies | Yes | Yes | Yes | Yes | Yes | Yes | Yes | Yes | Yes |
| Week Dummies | No | Yes | Yes | Yes | Yes | Yes | Yes | Yes | Yes |
| Observations | 1,100 | 1,100 | 1,100 | 1,100 | 1,100 | 1,100 | 1,100 | 1,100 | 1,100 |
| R-squared | 0.979 | 0.804 | 0.983 | 0.697 | 0.961 | 0.710 | 0.495 | 1.000 | 0.664 |

**2b.** **Y = Behavior $PH_{t-1}$→$B_{t-1}$**

| X = $PH_{t-1}$ | Transit Stations$_{t-1}$ | Retail & Recreation$_{t-1}$ | Grocery & Pharmacy$_{t-1}$ | Workplaces$_{t-1}$ | Residential$_{t-1}$ | Parks$_{t-1}$ | Driving$_{t-1}$ |
|---|---|---|---|---|---|---|---|
| SAHO[2] | -0.013 | -0.017 | -0.023* | -0.013 | 0.016 | -0.022 | 0.001 |
| | (0.011) | (0.011) | (0.013) | (0.009) | (0.011) | (0.015) | (0.008) |
| Daycare | -0.037*** | -0.051*** | -0.034* | 0.004 | 0.040*** | -0.023 | -0.043*** |
| | (0.010) | (0.015) | (0.018) | (0.013) | (0.014) | (0.027) | (0.011) |
| Business | 0.009 | -0.001 | -0.001 | 0.002 | -0.004 | 0.003 | -0.002 |
| | (0.005) | (0.006) | (0.005) | (0.006) | (0.006) | (0.005) | (0.006) |
| Religious | 0.009* | -0.002 | -0.006 | 0.008 | -0.017** | 0.003 | -0.008 |
| | (0.005) | (0.008) | (0.007) | (0.006) | (0.009) | (0.006) | (0.006) |
| Mask | 0.000 | -0.006 | -0.015** | -0.005 | 0.007 | 0.003 | 0.015*** |
| | (0.004) | (0.005) | (0.006) | (0.005) | (0.005) | (0.006) | (0.005) |
| Restaurant | -0.014* | -0.019** | -0.018** | -0.004 | 0.011 | -0.031*** | -0.017*** |
| | (0.008) | (0.008) | (0.009) | (0.006) | (0.007) | (0.009) | (0.006) |
| UI Ends[3] | -0.003 | 0.007 | -0.029*** | 0.011 | 0.003 | 0.013 | 0.038*** |
| | (0.007) | (0.008) | (0.010) | (0.007) | (0.007) | (0.009) | (0.008) |
| Tele Rx | 0.001 | 0.003 | 0.009* | 0.002 | -0.002 | -0.001 | -0.000 |
| | (0.004) | (0.005) | (0.005) | (0.004) | (0.004) | (0.004) | (0.004) |
| Constant | Yes | Yes | Yes | Yes | Yes | Yes | Yes |
| State Dummies | Yes | Yes | Yes | Yes | Yes | Yes | Yes |
| Week Dummies | Yes | Yes | Yes | Yes | Yes | Yes | Yes |
| Observations | 1,150 | 1,150 | 1,150 | 1,150 | 1,150 | 1,150 | 1,150 |
| R-squared | 0.839 | 0.843 | 0.595 | 0.918 | 0.896 | 0.569 | 0.743 |

| | |
|---|---|
| $\beta_1 = \sum \beta_{PH}$ | **2.831** |
| $\beta_2 = \sum \beta_I$ | **0.52** |
| $\beta_3 = \sum \beta_{PH,B}$ | **-0.26** |

Source: Own estimations using Household Pulse Survey

Robust standard errors in parentheses and significance levels*** p<0.01, ** p<0.05, * p<0.1

[1]Behavior variables come from google mobility data that measures traffic over the pandemic, these variables are standardized using min-max scaling $x_{scaled} = (x - x_{min})/(x_{max} - x_{min})$.

[2]SAHO – stay at home order by the week of pandemic 1= yes, 0 = no.

[3]Has the state ended unemployment insurance

*Interpretation: per unit change in covariate the coefficient ($\beta \times 100$) is the percent change.

**Table 3: Aggregate effect of case information$_{t-1}$, behavior$_{t-1}$ on economic distress by state.**

| | Y = Economic Distress $I_{t-1}\rightarrow ED_{t-1}$, $B_{t-1}\rightarrow ED_{t-1}$ (results of long regression) | | | |
|---|---|---|---|---|
| Dependent variable→ | **Log Cases$_{t-1}$** | **Job loss[3]$_{t-1}$** | **%Food Insecure2[4]** | **%Housing Insecure[5]** |
| 3a. Case Information $I_{t-1}\rightarrow ED_{t-1}$ | | | | |
| Log Cases$_{t-1}$[1] | | 0.041*** | 0.001 | 0.021** |
| | | (0.014) | (0.006) | (0.008) |
| Log Cases$_{t-2}$[2] | 0.520*** | | | |
| | (0.043) | | | |
| 3b. Behavior[6] $B_{t-1}\rightarrow ED_{t-1}$, and $B_{t-1}\rightarrow I_{t-1}$ | | | | |
| Transit Stations$_{t-1}$ | 0.041 | 0.001 | 0.009 | -0.001 |
| | (0.071) | (0.021) | (0.011) | (0.019) |
| Retail & Recreation$_{t-1}$ | -0.153** | -0.004 | -0.017 | -0.044* |
| | (0.073) | (0.027) | (0.015) | (0.023) |
| Grocery & Pharmacy$_{t-1}$ | 0.018 | -0.006 | 0.010 | 0.051*** |
| | (0.034) | (0.014) | (0.008) | (0.012) |
| Residential$_{t-1}$ | 0.042 | -0.021 | 0.019* | -0.003 |
| | (0.057) | (0.022) | (0.011) | (0.021) |
| Workplaces$_{t-1}$ | -0.068 | 0.025 | 0.020 | 0.024 |
| | (0.048) | (0.030) | (0.017) | (0.028) |
| Parks$_{t-1}$ | 0.007 | 0.026* | 0.009 | 0.009 |
| | (0.037) | (0.014) | (0.007) | (0.011) |
| Driving$_{t-1}$ | 0.008 | -0.001 | -0.012* | -0.014 |
| | (0.024) | (0.013) | (0.007) | (0.012) |
| $\beta_3$ | **0.52** | | | |
| $\beta_4 = \sum \beta_{4i}$ | **-0.105** | | | |
| $\beta_5 = \sum \beta_{5i}$ | **0.063** | | | |
| $\beta_6 = \sum \beta_{6b}$ | **0.08** | | | |
| Constant | Yes | Yes | Yes | Yes |
| Week | Yes | Yes | Yes | Yes |
| State Dummies | Yes | Yes | Yes | Yes |
| Observations | 1,100 | 1,150 | 1,150 | 1,150 |
| R-squared | 0.979 | 0.883 | 0.814 | 0.839 |

Source: Author analysis with Household Pulse Survey data
Robust standard errors in parentheses
*** p<0.01, ** p<0.05, * p<0.1
[1]In the equation where $I_{t-1}$ is the dependent variable $I_{t-2}$ is the independent covariate.
[2]$I_{t-1}$is the independent covariate.
[3]Job loss—is workers losing their work during the COVID-19 pandemic (temporary) and not perfectly correlated with unemployment rates.
[4]Food Insecure—respondents answered they did not have enough to eat or sometimes did not have enough to eat in past seven days.[5]Housing Insecure—not confident can pay mortgage or rent
[6]Food Insecure 2 in our calculations because only less than 6 percent experienced real food shortages.
*Interpretation: per unit change in covariate the coefficient ($\beta \times 100$) is the percent change.

## Table 4: Aggregate effect of information$_{t-1}$, public policies$_{t-1}$, economic distress on mental distress

| | Y = MD (results of short and long regression) | | | | | | | | |
|---|---|---|---|---|---|---|---|---|---|
| Dependent variable→ | MD[1] Continuous | | | GAD[2] | | | PHQ[3] | | |
| **Case Information $I_{t-2}$→$MD_t$, $I_{t-1}$→$MD_t$** | **Short** | **Long 1** | **Long 2** | **Short** | **Long 1** | **Long 2** | **Short** | **Long 1** | **Long 2** |
| Log Cases$_{t-1}$ | 0.065** | 0.118*** | 0.119** | 0.092*** | 0.156*** | 0.151*** | 0.037 | 0.081 | 0.086* |
| | (0.027) | (0.045) | (0.046) | (0.026) | (0.045) | (0.046) | (0.030) | (0.050) | (0.052) |
| Log Cases$_{t-2}$ | | -0.061** | -0.063** | | -0.055** | -0.055* | | -0.068** | -0.072** |
| | | (0.029) | (0.030) | | (0.028) | (0.029) | | (0.033) | (0.034) |
| **Economic Distress: $ED_{t-1}$→$MD_t$** | | | | | | | | | |
| Job loss$_{t-1}$[4] | 0.302*** | 0.289*** | 0.291*** | 0.276*** | 0.272*** | 0.273*** | 0.329*** | 0.306*** | 0.310*** |
| | (0.076) | (0.077) | (0.078) | (0.073) | (0.075) | (0.075) | (0.087) | (0.087) | (0.088) |
| Food Insecure$_{t-1}$[5] | 0.015 | -0.042 | -0.049 | 0.026 | -0.031 | -0.047 | 0.005 | -0.052 | -0.052 |
| | (0.146) | (0.148) | (0.148) | (0.139) | (0.142) | (0.141) | (0.168) | (0.169) | (0.170) |
| Housing Insecure$_{t-1}$[7] | 0.356*** | 0.390*** | 0.327** | 0.366*** | 0.386*** | 0.333*** | 0.345** | 0.394*** | 0.320** |
| | (0.124) | (0.128) | (0.130) | (0.118) | (0.122) | (0.124) | (0.141) | (0.146) | (0.148) |
| **Behavior[8] $B_{t-1}$→$MD_t$** | | | | | | | | | |
| Transit Stations$_{t-1}$ | | | 0.023 | | | 0.048 | | | -0.002 |
| | | | (0.045) | | | (0.041) | | | (0.052) |
| Retail & Recreation$_{t-1}$ | | | -0.062 | | | -0.077 | | | -0.047 |
| | | | (0.098) | | | (0.089) | | | (0.111) |
| Grocery & Pharmacy$_{t-1}$ | | | 0.108* | | | 0.080 | | | 0.136** |
| | | | (0.055) | | | (0.050) | | | (0.063) |
| Residential$_{t-1}$ | | | 0.032 | | | 0.062 | | | 0.002 |
| | | | (0.055) | | | (0.046) | | | (0.069) |
| Workplaces$_{t-1}$ | | | 0.119 | | | 0.132** | | | 0.106 |
| | | | (0.073) | | | (0.063) | | | (0.091) |
| Parks$_{t-1}$ | | | 0.011 | | | 0.023 | | | -0.001 |
| | | | (0.034) | | | (0.032) | | | (0.039) |
| Driving$_{t-1}$ | | | -0.015 | | | -0.024 | | | -0.006 |
| | | | (0.038) | | | (0.035) | | | (0.046) |
| $\beta_6 = \sum \beta_I$ | | **0.057** | **0.056** | | **0.101** | **0.096** | | **0.013** | **0.014** |
| $\beta_7 = \sum \beta_{EDi}$ | | **0.637** | **0.569** | | **0.627** | **0.559** | | **0.648** | **0.578** |
| $\beta_8 = \sum \beta_B$ | | | **0.216** | | | **0.244** | | | **0.188** |
| Constant | Yes | Yes | Yes | Yes | Yes | Yes | Yes | Yes | Yes |
| State Dummies | Yes | Yes | Yes | Yes | Yes | Yes | Yes | Yes | Yes |
| Week Dummies[9] | | | | | | | | | |
| Observations | 1,150 | 1,100 | 1,100 | 1,150 | 1,100 | 1,100 | 1,150 | 1,100 | 1,100 |
| R-squared | 0.632 | 0.661 | 0.674 | 0.703 | 0.710 | 0.722 | 0.595 | 0.600 | 0.619 |

Source: Own estimations using Household Pulse Survey

Robust standard errors in parentheses and significance levels*** p<0.01, ** p<0.05, * p<0.1

[1]MD—mental distress 4 items (anxious, worry, down , and lost interest) at time t

[2]GAD—Generalized anxiety disorder 2-item scale (feeling anxious or worried) at time t

[3]PHQ—Depression disorder 2-time scale (feeling down or lost interest) at time t

[4]Work loss—have not worked in past seven days

[5]Food Insecure 2: respondents answered they did not have enough to eat or sometimes did not have enough to eat in past seven days.

[6]Food Insecure—this is a perception of confidence in securing food over the next few weeks.

[7]Housing Insecure—not confident can pay mortgage or rent

[8]Behavior variables come from google mobility data that measures traffic over the pandemic, these variables are standardized using min-max scaling $x_{scaled} = (x - x_{min})/(x_{max} - x_{min})$.

[9]COVID cases changed dramatically week to week dummies. However, some survey weeks didn't show too much variation, and so they were dropped.

*Interpretation: per unit change in covariate the coefficient ($\beta \times 100$) is the percent change.

## Table 5: Causal Path Coefficients with cases as input

| Relationship | Coefficient | Source Table | Conversion Formula | Path Coefficient Value |
|---|---|---|---|---|
| Public Health | | | | |
| $I_{t-2}$→$PH_{t-1}$ | g | Table 2 | = $\beta_1$ | 2.831 |
| $B_{t-1}$→$PH_{t-1}$ | a | Table 2 (sum mobility on public health) | = $\beta_2$ | -0.26 |
| Case Information | | | | |
| $I_{t-1}$→$I_{t-2}$ | h | Table 2 | = $\beta_3$ | 0.522 |
| $I_{t-1}$→$B_{t-1}$ | b | Table 3 | = $\beta_4$ | -0.105 |
| Economic Distress | | | | |
| $I_{t-1}$→$ED_{t-1}$ | d | Table 3 | = $\beta_5$ | 0.063 |
| $B_{t-1}$→$ED_{t-1}$ | c | Table 3 | = $\beta_6$ | 0.08 |
| Mental Distress | | | | |
| $I_{t-1+t-2}$→$MD_t$ | e | Table 4 (long 2) | = $\beta_7$ | 0.056 |
| $B_{t-1}$→$MD_t$ | | Table 4 (long 2) | = $\beta_8$ | 0.216 |
| $ED_{t-1}$→$MD_t$ | f | Table 4 (long 2) | = $\beta_9$ | 0.569 |
| Generalized Anxiety | | | | |
| $I_{t-1+t-2}$→$GAD_t$ | e | Table 4 (long 2) | = $\beta_7$ | 0.096 |
| $B_t$→$GAD_t$ | | Table 4 (long 2) | = $\beta_8$ | 0.244 |
| $ED_{t-1}$→$GAD_t$ | f | Table 4 (long 2) | = $\beta_9$ | 0.559 |
| Depression | | | | |
| $I_{t-1+t-2}$→$PHQ_t$ | e | Table 4 (long 2) | = $\beta_7$ | 0.014 |
| $B_{t-1}$→$PHQ_t$ | | Table 4 (long 2) | = $\beta_8$ | 0.188 |
| $ED_{t-1}$→$PHQ_t$ | f | Table 4 (long 2) | = $\beta_9$ | 0.578 |

Source: Own calculations using HPS data
[1]I—case information at times t-1, and t-2
[2]B—behavior using mobility data at time t-1
[3]PH– Public health at time t-1
[4]ED – Economic distress at time t-1
[5]MD—mental distress 4 items (anxious, worry, down , and lost interest) at time t
[6]GAD—Generalized anxiety disorder 2-item scale (feeling anxious or worried) at time t
[7]PHQ—Depression disorder 2-time scale (feeling down or lost interest) at time t
*Interpretation: per unit change in covariate the coefficient ($\beta \times 100$) is the percent change.

**Figure 1: The causal pathway diagram for Mental Distress and estimated long regressions (numbered in red dotted lines)**

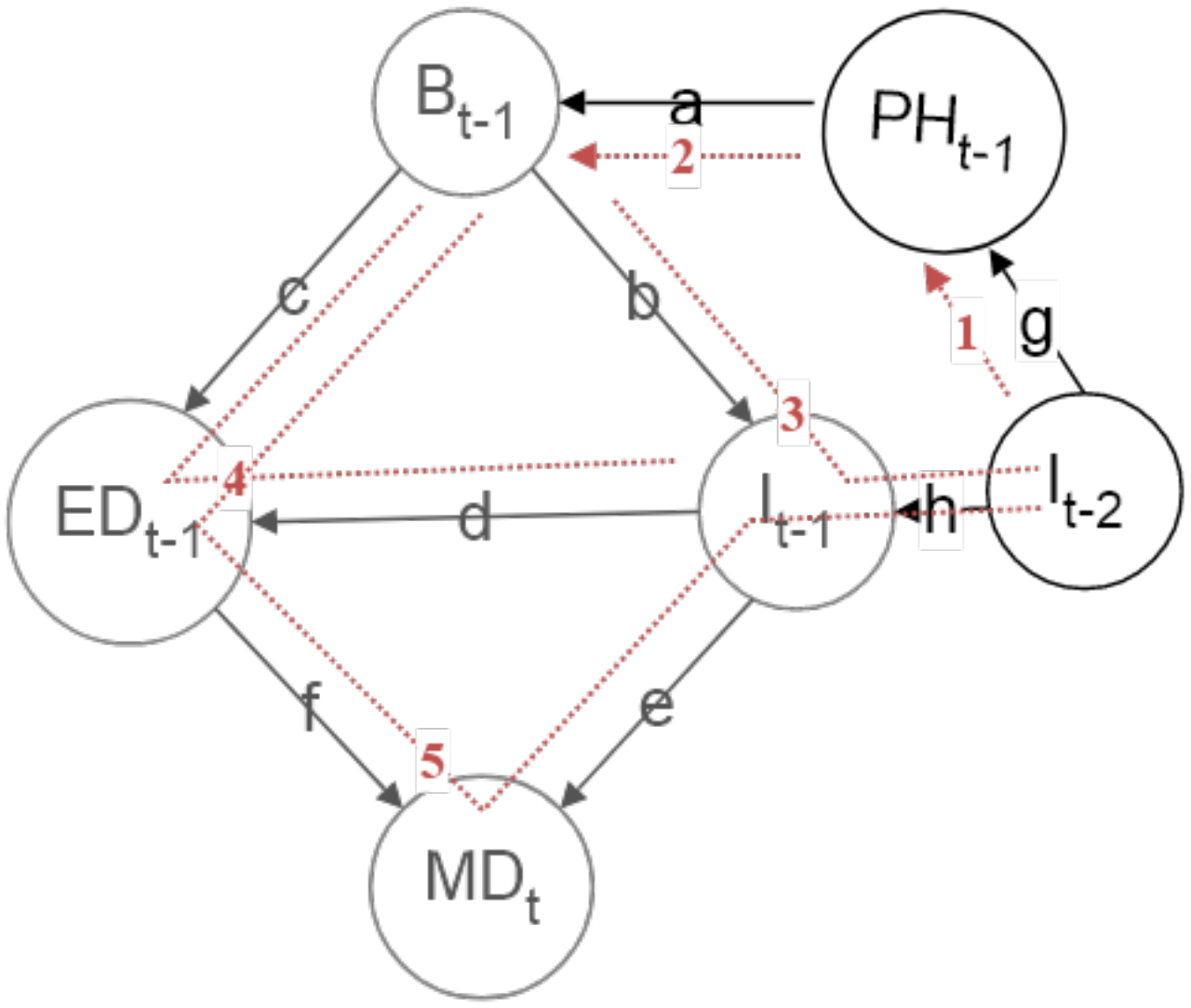

Note: path coefficients are the associated letters.

## Figure 2: Average Mental Distress in 50 U.S. States from April 23rd 2020 – February 15th 2021

2A. Proportion in Mental Distress by State and Week

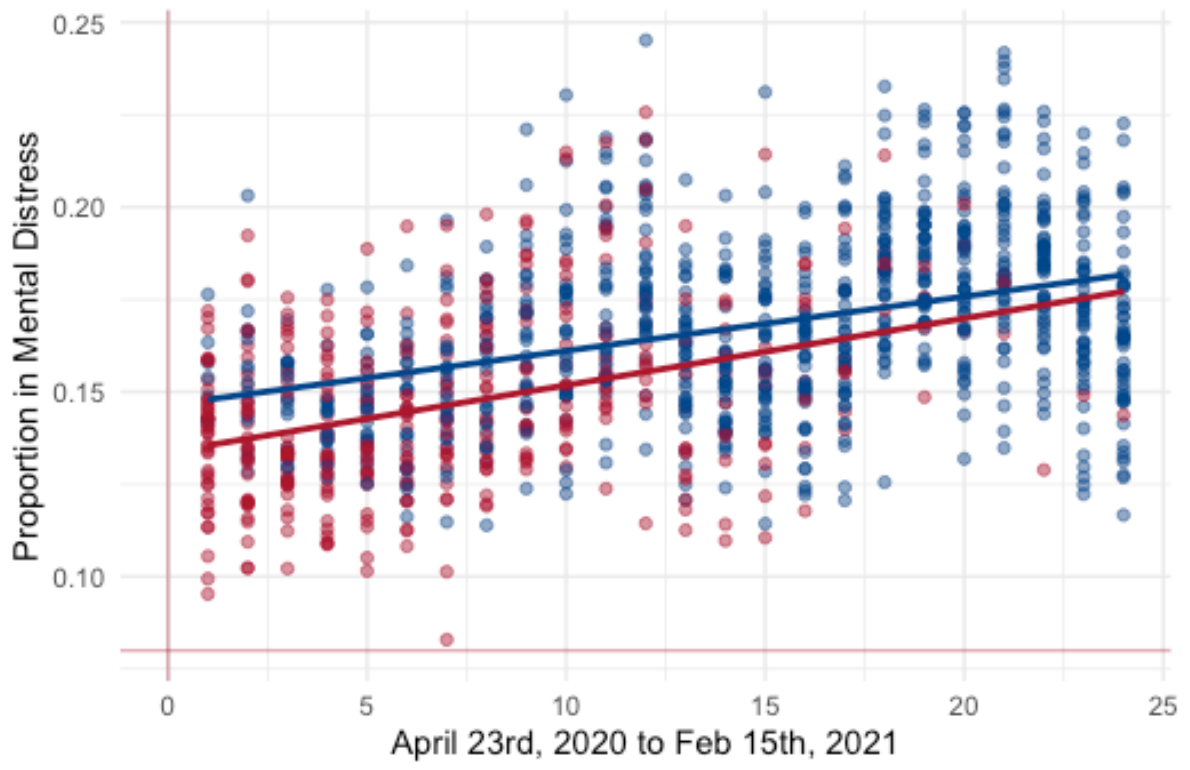

2B. NHIS proportion in Mental Distress 1st to 4th Quarter

A.

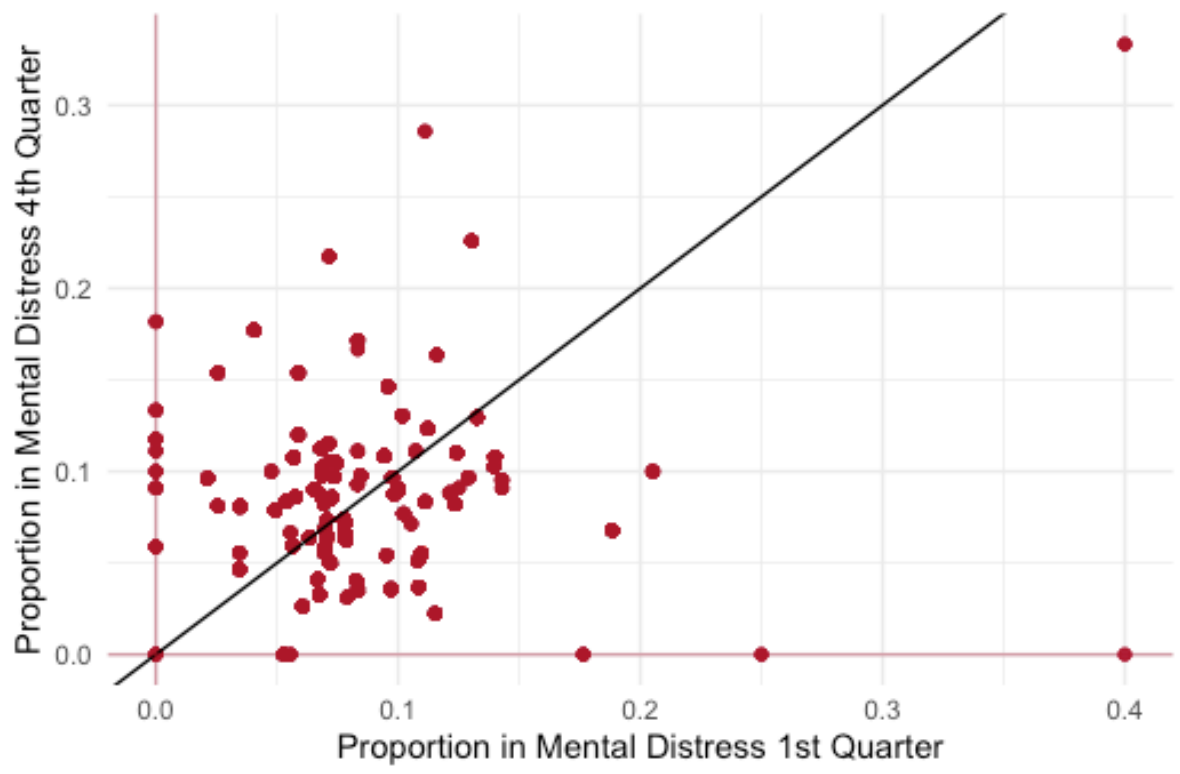

2C. Proportion in Mental Distress (half or more days)

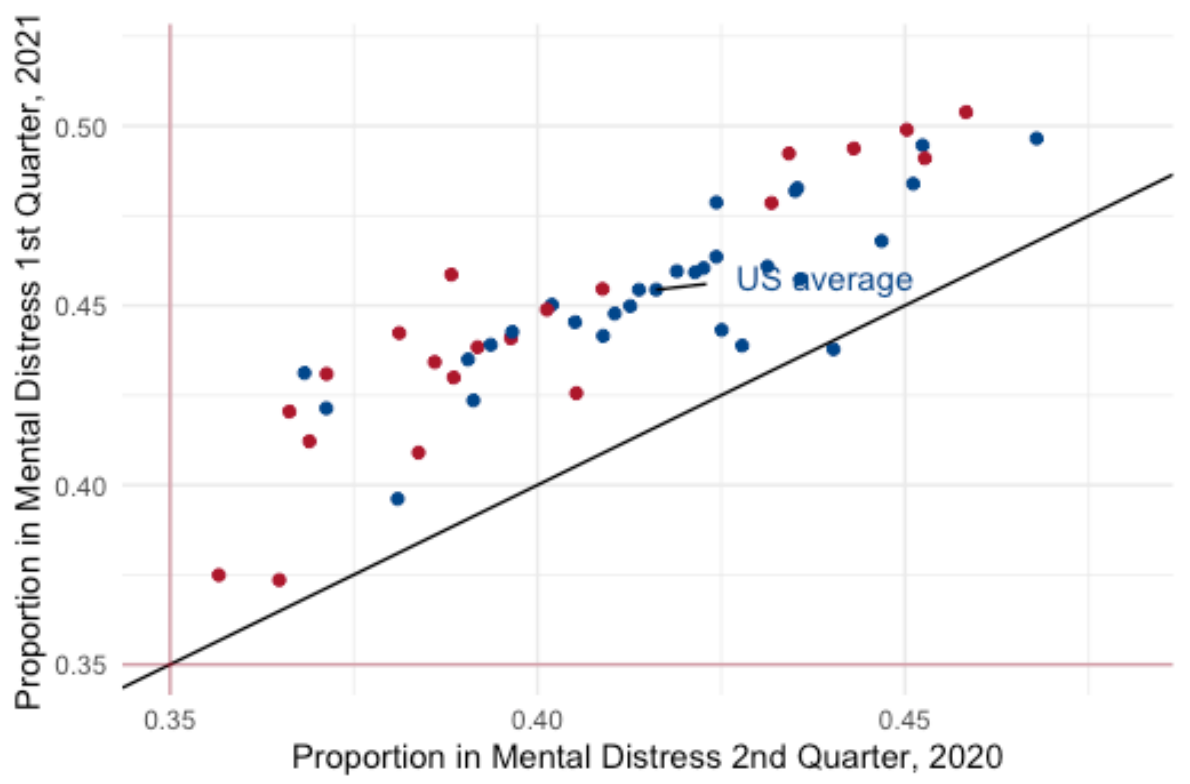

2D. Proportion in Mental Distress (nearly everyday)

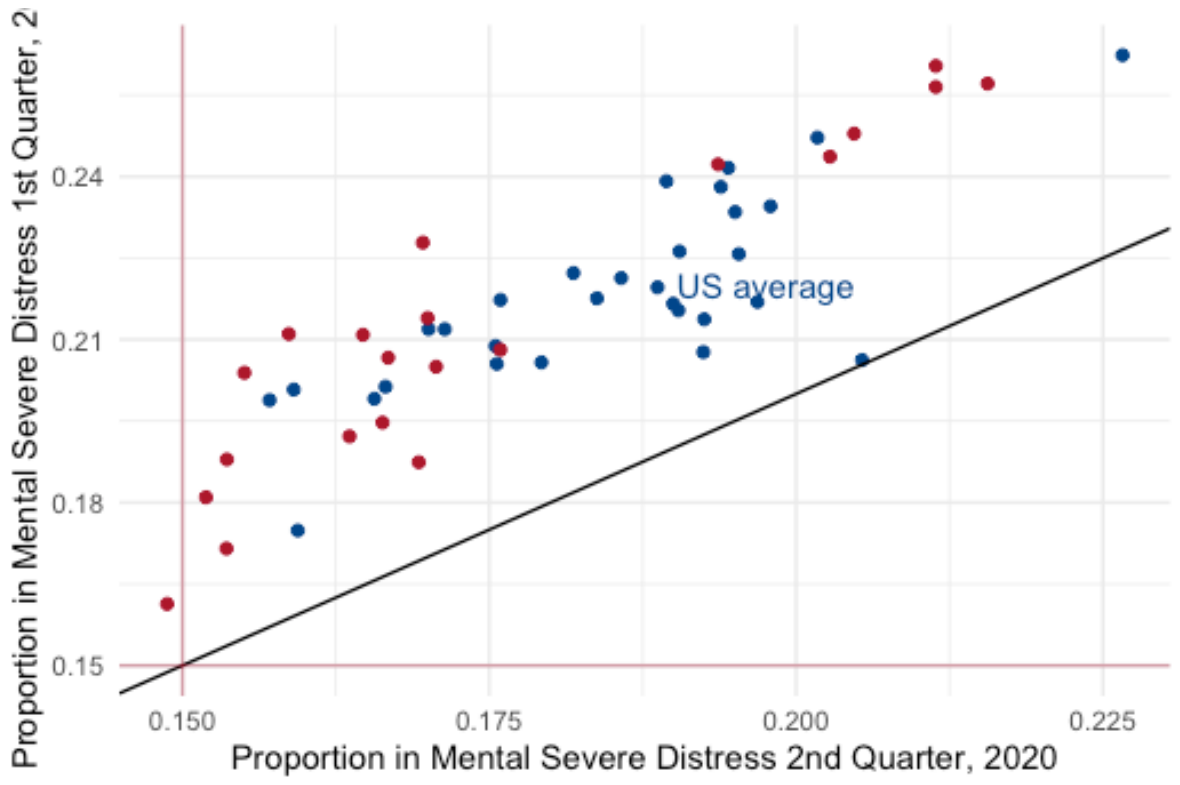

Proportion in mental distress by state for HPS 2020-2021. B. Proportion in mental distress 1st Quarter versus 4th Quarter by PSU for NHIS-2019. C. Proportion in mental distress for few days in the past 7 days 2nd Quarter 2020 versus 1st Quarter 2021 HPS 2020 – 2021. D. Proportion in mental distress for over half the days in the past 7 days 2nd Quarter 2020 versus 1st Quarter HPS 2021.

Note: Mental Distress is a composite index formulated by arithmetic mean values of two-item generalized anxiety disorder and two-item depressive disorder. The anxiety and depression scale responses took on the value 1 if there were zero days in these states, 2 if there were some bad days, 3 if over half the days were bad, and 4, if nearly every day was bad. For these graphs, we exclude those who report having no bad days during the survey period: Mental Distress = 1.

Note: NHIS does not have state-level data, so we use PSUs because they are the closest unit to instrument for state.

**Supplementary Materials**

## Mathematical Representation of DAG eFigure 1. Panel 2b

The assumptions underlying this directed acyclic graph are explained mathematically through a system of linear equations. A DAG is particularly useful in this estimation because we want to separate causes of pandemic related mental distress. We do assume a linear model for this estimation because the analysis is fairly straightforward. The sequence of arrows that lead to our primary variable of interest, mental health in the DAG, is assigned an “alphabet” these signify path coefficients that need to be recovered by estimating a system of linear equations and solving for them. The assumptions are as follows:

1. Mental health at time $t$ is determined after all other covariates at $\mathrm{t}-1$ are determined.
2. Public health policy $PH_{t-1}$ is determined by information, $\mathrm{I}_{\mathrm{t}-2}$,
3. Behavior $\mathrm{B}_{\mathrm{t}-1}$ is determined after public health policy $\mathrm{PH}_{\mathrm{t}-1}$ is determined at $\mathrm{t}-1$
4. Economic distress in $ED_{t-1}$ is determined by behavior $B_{t-1}$ and information $I_{t-1}$.
5. Mental Distress $MD_t$ is determined by economic distress $ED_{t-1}$ and information $I_{t-1}$.

From the above we can write down the mathematical equations:

$$\mathrm{PH}_{\mathrm{t}-1} = \propto_{\mathrm{PH}} + \beta_1 \mathrm{I}_{\mathrm{t}-2} + \mathrm{State.dummies} + \varepsilon_{\mathrm{PH}} \quad (1)$$
$$\mathrm{B}_{\mathrm{t}-1} = \propto_{\mathrm{B}} + \beta_2 \mathrm{PH}_{\mathrm{t}-1} + \mathrm{State.dummies} + \varepsilon_{\mathrm{B}} \quad (2)$$
$$\mathrm{I}_{\mathrm{t}-1} = \propto_{\mathrm{I}} + \beta_3 \mathrm{I}_{\mathrm{t}-2} + \beta_4 \mathrm{B}_{\mathrm{t}-1} + \mathrm{State.dummies} + \varepsilon_{\mathrm{I}} \quad (3)$$
$$\mathrm{ED}_{\mathrm{t}-1} = \propto_{\mathrm{ED}} + \beta_5 \mathrm{I}_{\mathrm{t}-1} + \beta_6 \mathrm{B}_{\mathrm{t}-1} + \mathrm{State.dummies} + \varepsilon_{\mathrm{ED}} \quad (4)$$
$$\mathrm{MD}_{\mathrm{t}} = \propto_{\mathrm{MD}} + \beta_7 \mathrm{I}_{\mathrm{t}-1} + \beta_8 \mathrm{I}_{\mathrm{t}-2} + \beta_9 \mathrm{ED}_{\mathrm{t}-1} + \mathrm{State.dummies} + \varepsilon_{\mathrm{MD}} \quad (5)$$

From equations (1-5) we can derive the path coefficients

i. $\beta_1 = g$
ii. $\beta_2 = a$
iii. $\beta_3 = \mathrm{h+gab}$
iv. $\beta_4 = b$
v. $\beta_5 = d$
vi. $\beta_6 = \mathrm{c}$
vii. $\beta_7 = e$
viii. $\beta_8 = eh$
ix. $\beta_9 = f$

We estimate mostly long regression in this case because there is a tradeoff between having too many backdoor relationships and the efficiency of long regressions. Tables e5-e7 show the correlations. The most important equation 5 we use both $\mathrm{I}_{\mathrm{t}-2}$ and $\mathrm{I}_{\mathrm{t}-1}$ because of a high degree of correlation, and if we argue that mental distress lingers, then it is plausible that $\mathrm{I}_{\mathrm{t}-2}$ has a causal relationship with $\mathrm{MD}_{\mathrm{t}}$.

## Appendix Estimated Models.

### eTable1: Summary of Aggregate (50 states) Variables Used (N = 1200, $N_{t-1}$=1150)

| Covariate | Description |
|---|---|
| Log Cases | Min-max scaled weekly cases |
| Log Deaths | Min-max scaled weekly deaths |
| **Behavior** | |
| Transit Stations$_{t-1}$ | |
| Retail & Recreation$_{t-1}$ | |
| Grocery & Pharmacy$_{t-1}$[5] | Min-max scaled weekly mobility data google |
| Workplaces$_{t-1}$ | |
| Parks$_{t-1}$ | |
| Driving$_{t-1}$ | Min-max scaled weekly mobility data apple |
| **Public Health Policies** | |
| **Binary per week** | |
| SAHO | Safe at home order |
| Daycare | Daycare closures |
| Restaurants | Restaurant closures |
| Religious | Religious gathering restrictions |
| Business | Business closures |
| Face Masks | Face Masks |
| UI Ends | Unemployment Insurance Ending |
| Tele Rx | Tele health and prescriptions |
| **Economic Distress Proportions by state** | |
| **Food Insecure 2** Proportion in the state that responded some times or often not enough to eat | In the last 7 days, which of these statements best describes the food eaten in your household? |
| **Job Loss** Proportion in the state that responded, yes. | Have you, or has anyone in your household experienced a loss of employment income since March 13, 2020? |
| **Housing Insecure** Proportion in the state that responded, not confident | How confident are you that your household will be able to pay your next rent or mortgage payment on time? |

## eTable 2: Correlation and scale validity for NHIS 2019 and HPS 2020

**Panel A: Pre-Covid Baseline: 4-item scale for Mental Distress (validity and correlations)-NHIS-2019, N = 31,258**

| | Dimension 1 | Dimension 2 | Dimension 3 | Dimension 4 |
|---|---|---|---|---|
| **Anxious**[1] | - | | | |
| **Worry**[1] | 0.69*** | - | | |
| **Down**[1] | 0.50*** | 0.50*** | - | |
| **Interest**[1] | 0.59*** | 0.60*** | 0.65*** | - |
| Mean | 1.35 | 1.28 | 1.26 | 1.25 |
| Standard Deviation | 0.72 | 0.71 | 0.67 | 0.63 |
| Interitem correlation | 0.58 | 0.58 | 0.63 | 0.56 |
| Range | 1-4 | 1-4 | 1-4 | 1-4 |
| Cronbach Alpha | 0.81 | 0.81 | 0.83 | 0.79 |
| Scale Test Mental Distress = 0.59 | | | | Alpha = 0.91 |

**Panel B: Pre-Covid Baseline correlation to actual diagnosis of Anxiety and Depression-NHIS 2019, N = 31, 258**

| | DSM Anxiety | DSM Depression | Mental Distress | % Formally with diagnoses and Mental Distress ≥ 2[4] |
|---|---|---|---|---|
| **DSM Anxiety**[2] | - | | | 55% |
| **DSM Depression**[2] | 0.47*** | - | | 62% |
| **Mental Distress**[3] | 0.50*** | 0.59*** | - | |

**Panel C: Post-Covid Household Pulse Survey 4-item scale for Mental Distress (validity and correlations)—HPS 2020, N = 1,650,188**

| | Dimension 1 | Dimension 2 | Dimension 3 | Dimension 4 |
|---|---|---|---|---|
| **Anxious**[1] | 1 | | | |
| **Worry**[1] | 0.80*** | 1 | | |
| **Down**[1] | 0.69*** | 0.72*** | 1 | |
| **Interest**[1] | 0.63*** | 0.67*** | 0.78*** | 1 |
| Mean | 2.14 | 1.95 | 1.86 | 1.89 |
| Standard Deviation | 1.08 | 1.04 | 1.00 | 1.00 |
| Interitem correlation | 0.72 | 0.70 | 0.70 | 0.74 |
| Range | 1-4 | 1-4 | 1-4 | 1-4 |
| Cronbach Alpha | 0.89 | 0.87 | 0.88 | 0.89 |
| Scale Test = 0.71 | | | | Alpha = 0.91 |

Source: Own calculations using NHIS and HPS publicly available data
NHIS: National Health Interview Survey, 2019
HPS: Household Pulse Survey-weekly, 2020
[1]All four dimensions anxiety, worry, feeling down, and losing interest appear in both surveys. They take on the value 1 = no days, 2 = some days, 3 = over half days, and 4 = nearly every day.
[2]DSM anxiety and DSM depression are clinical diagnoses of generalized anxiety disorder and depressive disorder.
[3]Mental Distress is calculated as the arithmetic mean of the four dimensions anxiety, worry, down, and interest.
[4]Baseline Mental Distress ≥ 2 and formal diagnosis of anxiety disorder or depressive disorder

## eTable 3: Aggregate effects of information (log deaths)$_{t-2}$, behavior$_{t-1}$[1] on public policies$_{t-1}$ by state.

| 2a. | | | | | Y = Log deaths $I_{t-2} \rightarrow I_{t-1}$, Public Health Policy $I_{t-2} \rightarrow PH_{t-1}$ | | | | |
|---|---|---|---|---|---|---|---|---|---|
| Dependent Variable | Log deaths $I_{t-1}$ | SAHO[2] | Daycare | Business | Religious | Mask | Restaurant | UI ends | Tele Rx |
| Log deaths $I_{t-2}$ | 0.479*** | | | | | | | | |
| | (0.042) | | | | | | | | |
| | | -0.193 | -0.016 | -0.082 | 0.118 | 0.843*** | 0.685** | -0.000*** | 0.532* |
| | | (0.141) | (0.025) | (0.336) | (0.083) | (0.240) | (0.309) | (0.000) | (0.283) |
| Constant | Yes | Yes | Yes | Yes | Yes | Yes | Yes | Yes | Yes |
| State Dummies | Yes | Yes | Yes | Yes | Yes | Yes | Yes | Yes | Yes |
| Week Dummies | No | Yes | Yes | Yes | Yes | Yes | Yes | Yes | Yes |
| Observations | 1,100 | 1,100 | 1,100 | 1,100 | 1,100 | 1,100 | 1,100 | 1,100 | 1,100 |
| R-squared | 0.981 | 0.804 | 0.983 | 0.696 | 0.961 | 0.703 | 0.492 | 1.000 | 0.663 |
| 2b. | | | | | Y = Behavior $PH_{t-1} \rightarrow B_{t-1}$ | | | | |
| X = $PH_{t-1}$ | | | | | | | | | |
| SAHO[2] | | -0.013 | -0.017 | -0.023* | -0.013 | 0.016 | -0.022 | | 0.001 |
| | | (0.011) | (0.011) | (0.013) | (0.009) | (0.011) | (0.015) | | (0.008) |
| Daycare | | -0.037*** | -0.051*** | -0.034* | 0.004 | 0.040*** | -0.023 | | -0.043*** |
| | | (0.010) | (0.015) | (0.018) | (0.013) | (0.014) | (0.027) | | (0.011) |
| Business | | 0.009 | -0.001 | -0.001 | 0.002 | -0.004 | 0.003 | | -0.002 |
| | | (0.005) | (0.006) | (0.005) | (0.006) | (0.006) | (0.005) | | (0.006) |
| Religious | | 0.009* | -0.002 | -0.006 | 0.008 | -0.017** | 0.003 | | -0.008 |
| | | (0.005) | (0.008) | (0.007) | (0.006) | (0.009) | (0.006) | | (0.006) |
| Mask | | 0.000 | -0.006 | -0.015** | -0.005 | 0.007 | 0.003 | | 0.015*** |
| | | (0.004) | (0.005) | (0.006) | (0.005) | (0.005) | (0.006) | | (0.005) |
| Restaurant | | -0.014* | -0.019** | -0.018** | -0.004 | 0.011 | -0.031*** | | -0.017*** |
| | | (0.008) | (0.008) | (0.009) | (0.006) | (0.007) | (0.009) | | (0.006) |
| UI Ends[3] | | -0.003 | 0.007 | -0.029*** | 0.011 | 0.003 | 0.013 | | 0.038*** |
| | | (0.007) | (0.008) | (0.010) | (0.007) | (0.007) | (0.009) | | (0.008) |
| Tele Rx | | 0.001 | 0.003 | 0.009* | 0.002 | -0.002 | -0.001 | | -0.000 |
| | | (0.004) | (0.005) | (0.005) | (0.004) | (0.004) | (0.004) | | (0.004) |
| Constant | | Yes | Yes | Yes | Yes | Yes | Yes | | Yes |
| State Dummies | | Yes | Yes | Yes | Yes | Yes | Yes | | Yes |
| Week Dummies | | 1,150 | 1,150 | 1,150 | 1,150 | 1,150 | 1,150 | | 1,150 |
| Observations | | 0.839 | 0.843 | 0.595 | 0.918 | 0.896 | 0.569 | | 0.743 |
| R-squared | | | | | | | | | |
| $\beta_1 = \sum \beta_{PH}$ | **1.887** | | | | | | | | |
| $\beta_2 = \sum \beta_I$ | **0.479** | | | | | | | | |
| $\beta_3 = \sum \beta_{PH,B}$ | **-0.26** | | | | | | | | |

Source: Own estimations using Household Pulse Survey

Robust standard errors in parentheses and significance levels*** p<0.01, ** p<0.05, * p<0.1

[1]Behavior variables come form google mobility data that measures traffic over the pandemic, these variables are standardized using min-max scaling $x_{scaled} = (x - x_{min})/(x_{max} - x_{min})$.

[2]SAHO – stay at home order by week of pandemic 1= yes, 0 = no.

[3]Has the state ended unemployment insurance

## eTable 4: Aggregate effect of case information$_{t-1}$, behavior$_{t-1}$ on economic distress by state.

| | Y = Economic Distress $I_{t-1} \rightarrow ED_{t-1}$, $B_{t-1} \rightarrow ED_{t-1}$ (results of long regression) | | | |
|---|---|---|---|---|
| Dependent variable→ | **Log Deaths** | **Job loss** | **%Food Insecure [3]** | **%Housing Insecure[5]** |
| 3a. Case Information $I_{t-1} \rightarrow ED_{t-1}$ | | | | |
| Log Deaths$_{t-1}$[1] | | 0.048*** | 0.001 | 0.016* |
| | | (0.016) | (0.007) | (0.009) |
| Log Deaths$_{t-2}$[2] | 0.409*** | | | |
| | (0.036) | | | |
| 3b. Behavior[6] $B_{t-1} \rightarrow ED_{t-1}$, and $B_{t-1} \rightarrow I_{t-1}$ | | | | |
| Transit Stations$_{t-1}$ | 0.043 | 0.000 | 0.009 | -0.001 |
| | (0.056) | (0.021) | (0.011) | (0.019) |
| Retail & Recreation$_{t-1}$ | -0.043 | -0.008 | -0.017 | -0.046* |
| | (0.050) | (0.027) | (0.015) | (0.024) |
| Grocery & Pharmacy$_{t-1}$ | -0.038 | -0.004 | 0.010 | 0.051*** |
| | (0.027) | (0.014) | (0.008) | (0.013) |
| Residential$_{t-1}$ | -0.054 | -0.015 | 0.020* | -0.000 |
| | (0.050) | (0.023) | (0.011) | (0.022) |
| Workplaces$_{t-1}$ | -0.125*** | 0.029 | 0.020 | 0.025 |
| | (0.044) | (0.030) | (0.017) | (0.028) |
| Parks$_{t-1}$ | -0.001 | 0.026* | 0.009 | 0.010 |
| | (0.029) | (0.014) | (0.007) | (0.012) |
| Driving$_{t-1}$ | 0.008 | -0.000 | -0.012* | -0.013 |
| | (0.023) | (0.013) | (0.007) | (0.012) |
| $\beta_3$ | **0.409** | | | |
| $\beta_4 = \sum \beta_{4i}$ | **-0.21** | | | |
| $\beta_5 = \sum \beta_{5i}$ | **0.065** | | | |
| $\beta_6 = \sum \beta_{6b}$ | **-0.093** | | | |
| Constant | Yes | Yes | Yes | Yes |
| State Dummies | Yes | Yes | Yes | Yes |
| Observations | 1,100 | 1,150 | 1,150 | 1,150 |
| R-squared | 0.980 | 0.883 | 0.814 | 0.839 |

Source: Author analysis with Household Pulse Survey data
Robust standard errors in parentheses
*** p<0.01, ** p<0.05, * p<0.1
[1]In the equation where $I_{t-1}$ is the dependent variable $I_{t-2}$ is the independent covariate.
[2]$I_{t-1}$is the independent covariate.
[3]Food Insecure: respondents answered they did not have enough to eat or sometimes did not have enough to eat in past seven days.
[5]Housing Insecure—not confident can pay mortgage or rent
[6]Food Insecure 2 in our calculations because only less than 6 percent experienced real food shortages.

**eTable 5: Aggregate effect of COVID-19 case death information$_{t-1}$, public policies$_{t-1}$, economic distress on mental distress**

| | Y = MD (results of short and long regression) | | | | | | | | |
|---|---|---|---|---|---|---|---|---|---|
| Dependent variable→ | | MD[1] Continuous | | | GAD[2] | | | | PHQ[3] |
| **Case Information $I_{t-2}$→$MD_t$, $I_{t-1}$→$MD_t$** | **Short** | **Long 1** | **Long 2** | **Short** | **Long 1** | **Long 2** | **Short** | **Long 1** | **Long 2** |
| Log Deaths$_{-1}$ | -0.010 | -0.055 | -0.048 | 0.020 | -0.004 | -0.000 | -0.041 | -0.106* | -0.095 |
| | (0.032) | (0.055) | (0.056) | (0.031) | (0.054) | (0.055) | (0.037) | (0.060) | (0.061) |
| Log Deaths$_{t-2}$ | | -0.030 | -0.034 | | -0.033 | -0.035 | | -0.027 | -0.032 |
| | | (0.035) | (0.036) | | (0.034) | (0.035) | | (0.039) | (0.040) |
| **Economic Distress: $ED_{t-1}$→$MD_t$,** | | | | | | | | | |
| Job loss$_{t-1}$[8] | 0.325*** | 0.323*** | 0.321*** | 0.297*** | 0.306*** | 0.302*** | 0.352*** | 0.340*** | 0.341*** |
| | (0.077) | (0.078) | (0.078) | (0.074) | (0.076) | (0.076) | (0.088) | (0.088) | (0.088) |
| Food Insecure$_{t-1}$ | -0.001 | -0.070 | -0.075 | 0.010 | -0.060 | -0.074 | -0.011 | -0.079 | -0.077 |
| | (0.146) | (0.148) | (0.148) | (0.140) | (0.142) | (0.142) | (0.167) | (0.169) | (0.169) |
| Housing Insecure$_{t-1}$[10] | 0.369*** | 0.417*** | 0.354*** | 0.384*** | 0.422*** | 0.367*** | 0.354** | 0.413*** | 0.341** |
| | (0.124) | (0.128) | (0.130) | (0.119) | (0.123) | (0.125) | (0.141) | (0.146) | (0.148) |
| **Behavior $B_{t-1}$→$MD_t$** | | | | | | | | | |
| Transit Stations$_{t-1}$ | | | 0.033 | | | 0.058 | | | 0.008 |
| | | | (0.044) | | | (0.041) | | | (0.052) |
| Retail & Recreation$_{t-1}$ | | | -0.086 | | | -0.105 | | | -0.067 |
| | | | (0.100) | | | (0.094) | | | (0.111) |
| Grocery & Pharmacy$_{t-1}$[5] | | | 0.105* | | | 0.078 | | | 0.131** |
| | | | (0.057) | | | (0.052) | | | (0.064) |
| Residential$_{t-1}$ | | | 0.026 | | | 0.059 | | | -0.006 |
| | | | (0.053) | | | (0.045) | | | (0.067) |
| Workplaces$_{t-1}$ | | | 0.100 | | | 0.116* | | | 0.084 |
| | | | (0.072) | | | (0.063) | | | (0.089) |
| Parks$_{t-1}$ | | | 0.015 | | | 0.028 | | | 0.002 |
| | | | (0.034) | | | (0.032) | | | (0.039) |
| Driving$_{t-1}$ | | | -0.010 | | | -0.019 | | | -0.002 |
| | | | (0.038) | | | (0.035) | | | (0.045) |
| $\beta_6 = \sum \beta_I$ | | -0.085 | -0.082 | | -0.037 | -0.035 | | -0.133 | -0.127 |
| $\beta_7 = \sum \beta_{EDi}$ | | 0.67 | 0.6 | | 0.668 | 0.595 | | 0.674 | 0.605 |
| $\beta_8 = \sum \beta_B$ | | | 0.183 | | | 0.215 | | | 0.15 |
| Constant | Yes | Yes | Yes | Yes | Yes | Yes | Yes | Yes | Yes |
| State Dummies | Yes | Yes | Yes | Yes | Yes | Yes | Yes | Yes | Yes |
| Week Dummies | Yes | Yes | Yes | Yes | Yes | Yes | Yes | Yes | Yes |
| Observations | 1,050 | 1,000 | 1,000 | 1,050 | 1,000 | 1,000 | 1,050 | 1,000 | 1,000 |
| R-squared | 0.652 | 0.663 | 0.707 | 0.687 | 0.696 | 0.714 | 0.608 | 0.623 | 0.742 |

Source: Own estimations using Household Pulse Survey

Robust standard errors in parentheses and significance levels*** p<0.01, ** p<0.05, * p<0.1

[1]MD—mental distress 4 items (anxious, worry, down , and lost interest) at time t

[2]GAD—Generalized anxiety disorder 2-item scale (feeling anxious or worried) at time t

[3]PHQ—Depression disorder 2-time scale (feeling down or lost interest) at time t

[4]Work loss—have not worked in past seven days

[5]Food Insecure 2: respondents answered they did not have enough to eat or sometimes did not have enough to eat in past seven days.

[6]Food Insecure—this is a perception of confidence in securing food over the next few weeks.

[7]Housing Insecure—not confident can pay mortgage or rent

[8]Behavior variables come form google mobility data that measures traffic over the pandemic, these variables are standardized using min-max scaling $x_{scaled} = (x - x_{min})/(x_{max} - x_{min})$.

## eTable 6: Causal Path Coefficients with deaths as input

| Relationship | Coefficient | Source Table | Conversion Formula | Path Coefficient Value |
|---|---|---|---|---|
| Public Health | | | | |
| $I_{t-2} \rightarrow PH_{t-1}$ | g | Table 2 | $= \beta_1$ | 1.887 |
| $B_{t-1} \rightarrow PH_{t-1}$ | a | Table 2 (sum mobility) | $= \beta_2$ | -0.26 |
| Case Information | | | | |
| $I_{t-1} \rightarrow I_{t-2}$ | h | Table 2 | $= \beta_3$ | 0.479 |
| $I_{t-1} \rightarrow B_{t-1}$ | b | Table 3 | $= \beta_4$ | -0.21 |
| Economic Distress | | | | |
| $I_{t-1} \rightarrow ED_{t-1}$ | d | Table 3 | $= \beta_5$ | 0.065 |
| $B_{t-1} \rightarrow ED_{t-1}$ | c | Table 3 | $= \beta_6$ | 0.093 |
| Mental Distress | | | | |
| $I_{t-1} \rightarrow MD_t$ | e | Table 4 (long 2) | $= \beta_7$ | -0.048 |
| $I_{t-2} \rightarrow MD_t$ | eh | Table 4 (long 2) | $= \beta_8$ | -0.034 |
| $ED_{t-1} \rightarrow MD_t$ | f | Table 4 (long 2) | $= \beta_9$ | 0.6 |
| $B_{t-1} \rightarrow MD_t$ | | Table 4 (long 2) | | 0.18 |
| Generalized Anxiety | | | | |
| $I_{t-1} \rightarrow GAD_t$ | e | Table 4 (long 1) | $= \beta_7$ | -0.00 |
| $I_{t-2} \rightarrow GAD_t$ | eh | Table 4 (long 1) | $= \beta_8$ | -0.35 |
| $B_{t-1} \rightarrow MD_t$ | | Table 4 (long 2) | | 0.215 |
| $ED_{t-1} \rightarrow GAD_t$ | f | Table 4 (long 1) | $= \beta_9$ | 0.595 |
| Depression | | | | |
| $I_{t-1} \rightarrow PHQ_t$ | e | Table 4 (long 1) | $= \beta_7$ | -0.095 |
| $I_{t-2} \rightarrow PHQ_t$ | eh | Table 4 (long 1) | $= \beta_8$ | -0.032 |
| $B_{t-1} \rightarrow MD_t$ | | Table 4 (long 2) | | 0.15 |
| $ED_{t-1} \rightarrow PHQ_t$ | f | Table 4 (long 1) | $= \beta_9$ | 0.605 |

Source: Own calculations using HPS data
[1]I—case information at times t-1, and t-2
[2]B—behavior using mobility data at time t-1
[3]PH– Public health at time t-1
[4]ED – Economic distress at time t-1
[5]MD—mental distress 4 items (anxious, worry, down , and lost interest) at time t
[6]GAD—Generalized anxiety disorder 2-item scale (feeling anxious or worried) at time t
[7]PHQ—Depression disorder 2-time scale (feeling down or lost interest) at time t

## eFigure 1: Causal pathways to Mental Distress showing backdoor relationships on short regressions

Panel 1: $PH_{t-1}$, $I_{t-1}$, and $B_{t-1}$ get determined first from $I_{t-2}$ cases.

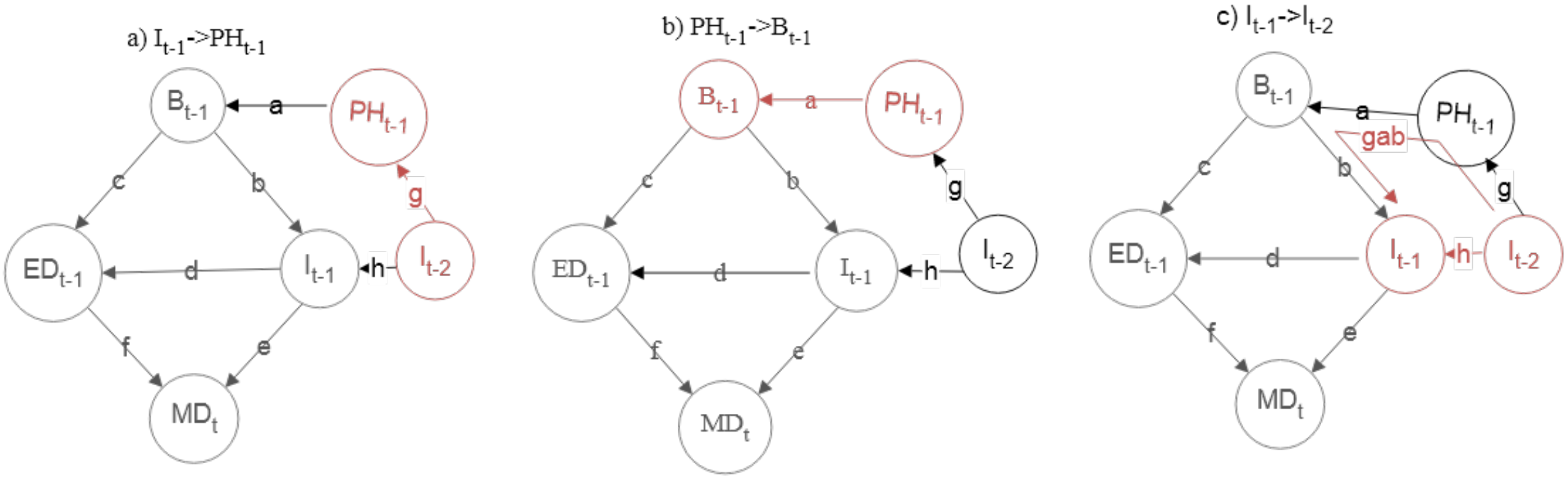

Panel 2: Given $I_{t-2}$, $PH_{t-1}$, $I_{t-1}$, and $B_{t-1}$: Current impact on $ED_{t-1}$ and future impact on $MD_t$ get determined.

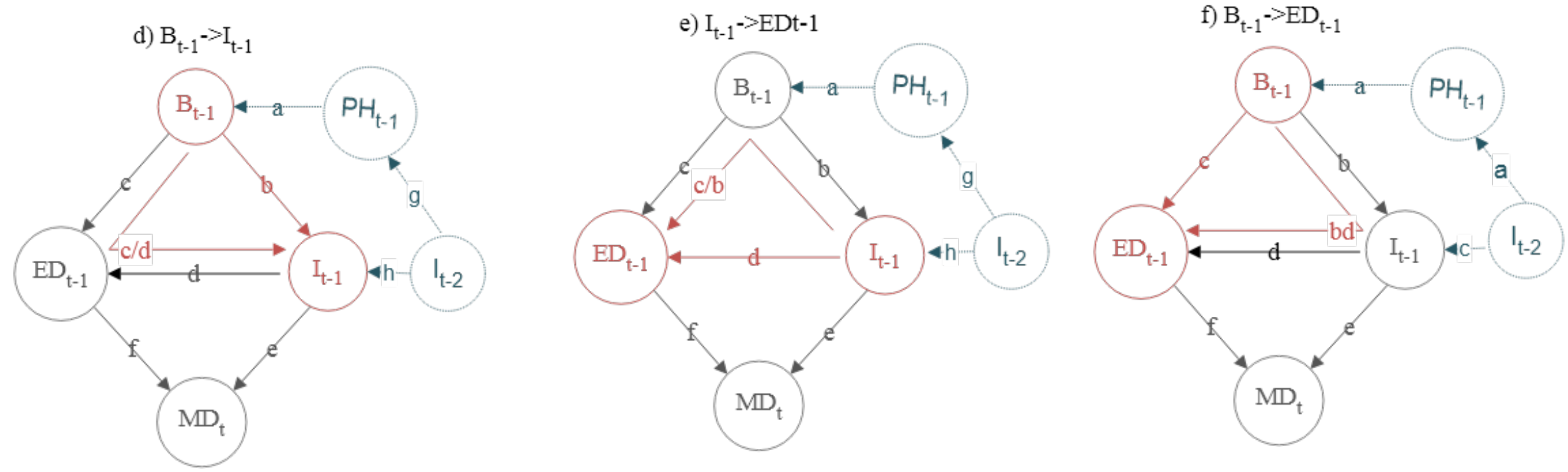

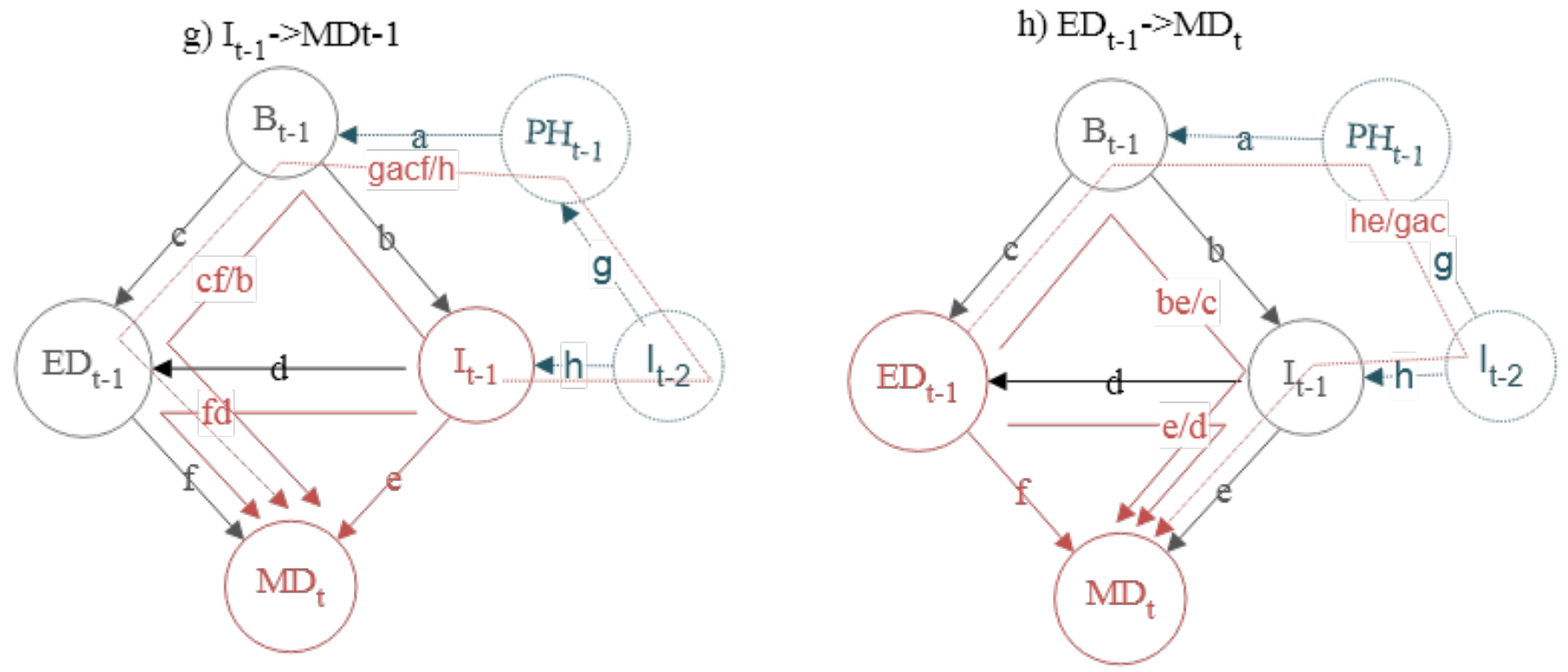

For $MD_t$ because of high correlation with $I_{t-2}$ we permit secondary effects. This means for obtaining accurate path coefficients we must in the long regression include all covariates that could influence $MD_t$

**eFigure 2: Proportions in Economic Distress (Job loss, food insecure or housing insecure) versus proportions in Mental Distress April 23rd – June 30th 2020 versus January 01 – February 15th 2021**

e2A. Proportion in Job-loss vs. Mental Distress

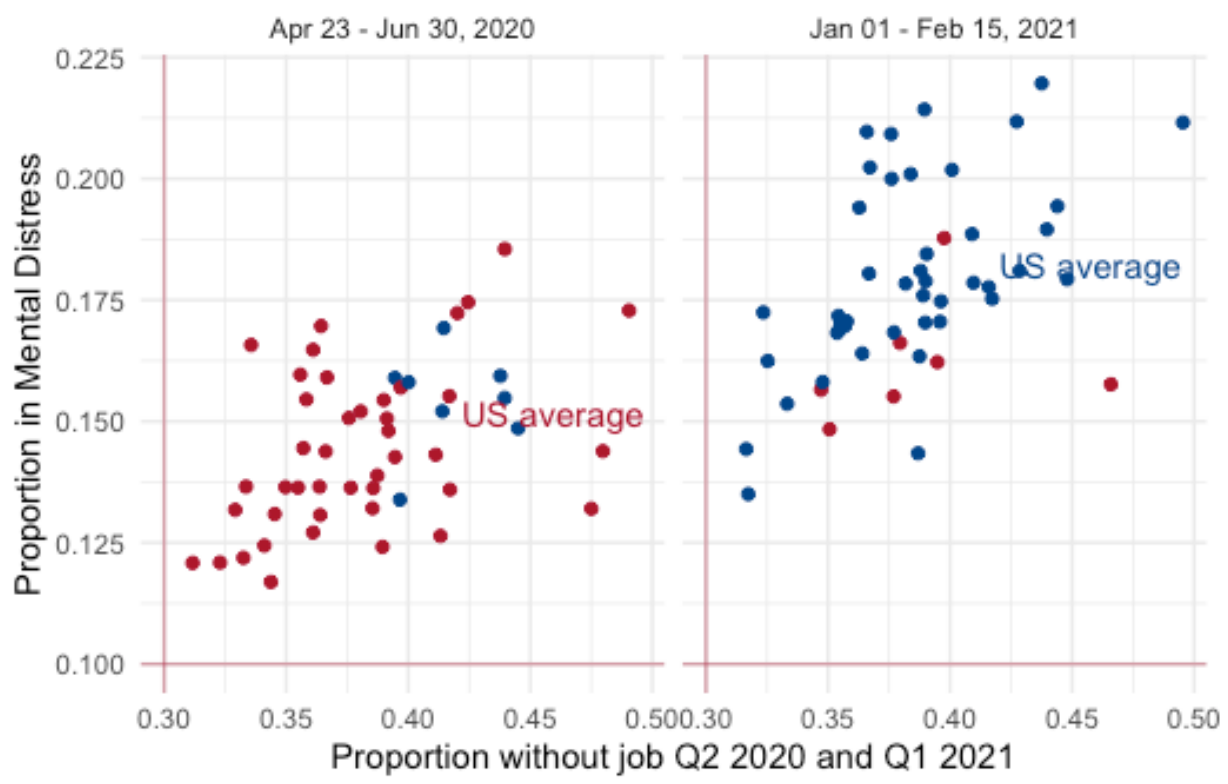

e2B. Proportion Housing Insecure vs. Mental Distress

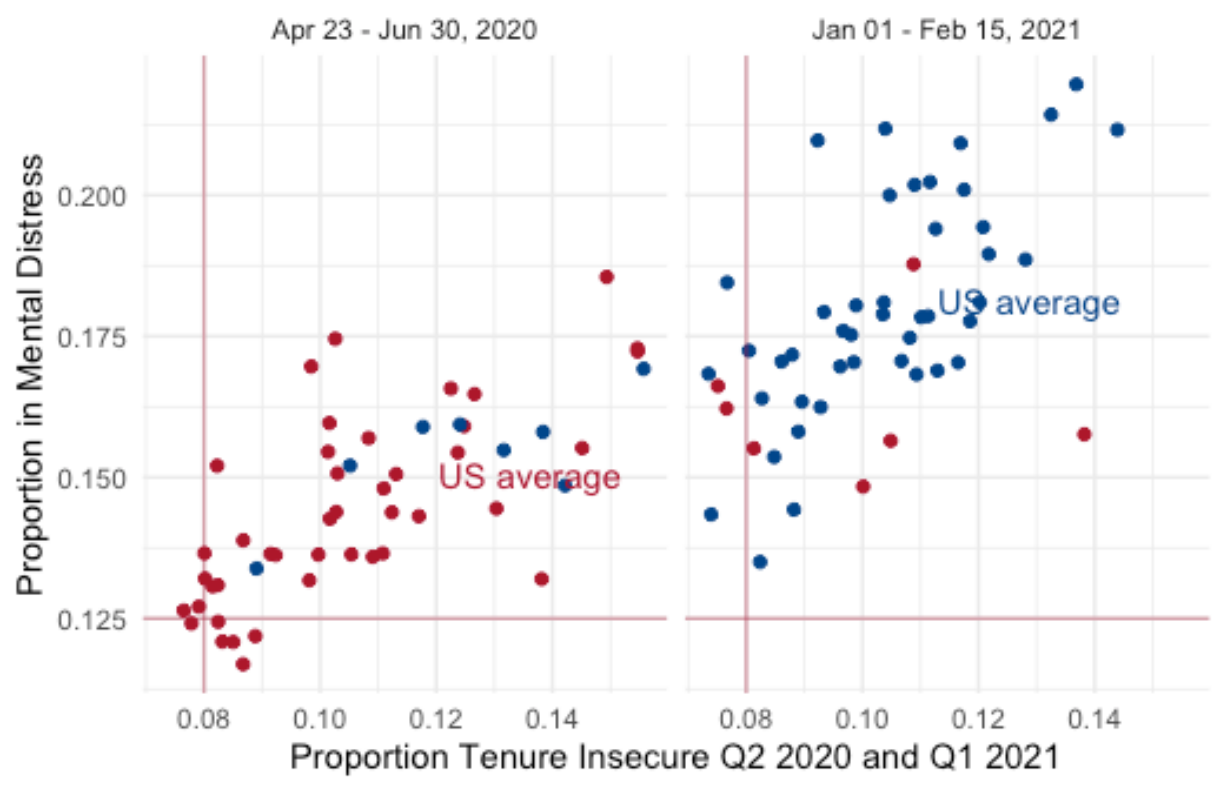

e2C. Proportion Food Insecure vs. Mental Distress

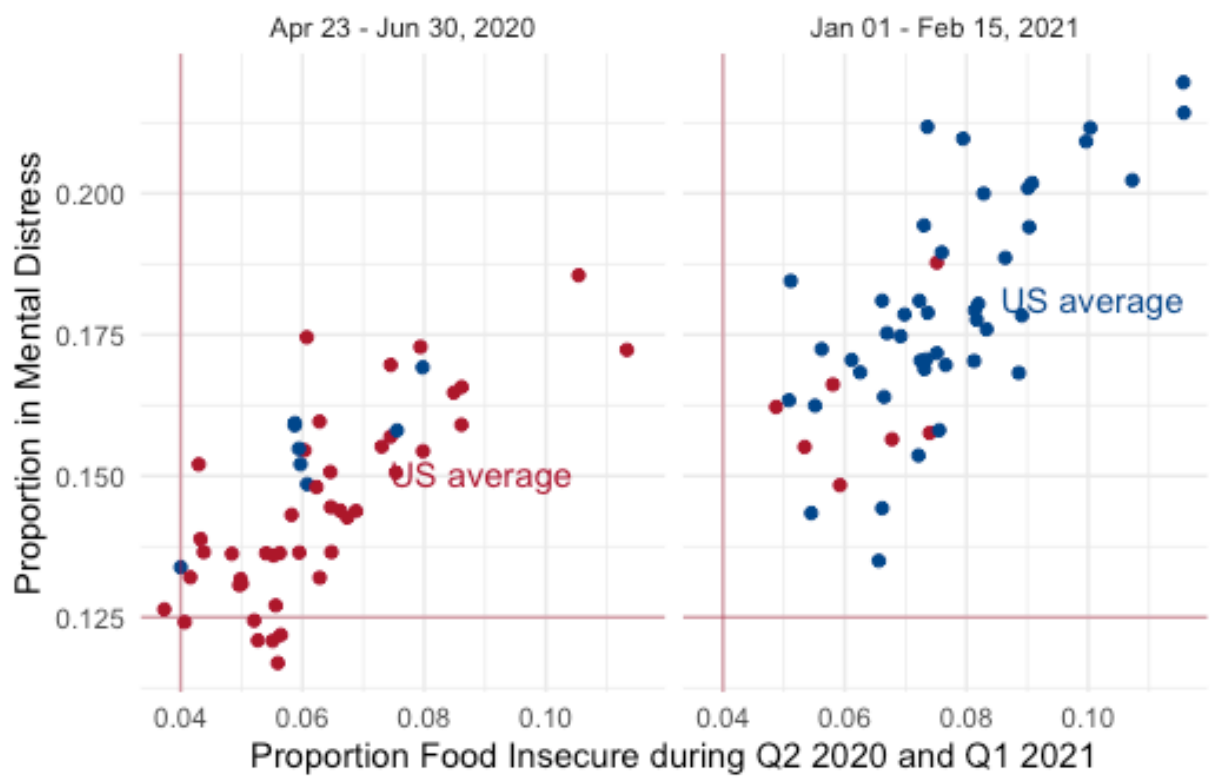

eFigure 1: Proportion in economic distress and proportion in mental distress during 2nd Quarter 2020 and 1st Quarter 2021.
e1A. The proportion in mental distress with pandemic related job-loss during 2nd Quarter 2020 and 1st Quarter 2021.
e1B. The proportion in mental distress with pandemic related food insecurity during 2nd Quarter 2020 and 1st Quarter 2021.
e1C. The proportion in mental distress with pandemic related housing insecurity during 2nd Quarter 2020 and 1st Quarter 2021.

**eFigure 3. Behavior during pandemic using min-max scaled aggregated mobility patterns by state from April 23rd 2021 – February 15th 2021.**

e3A. Apple Driving Patterns

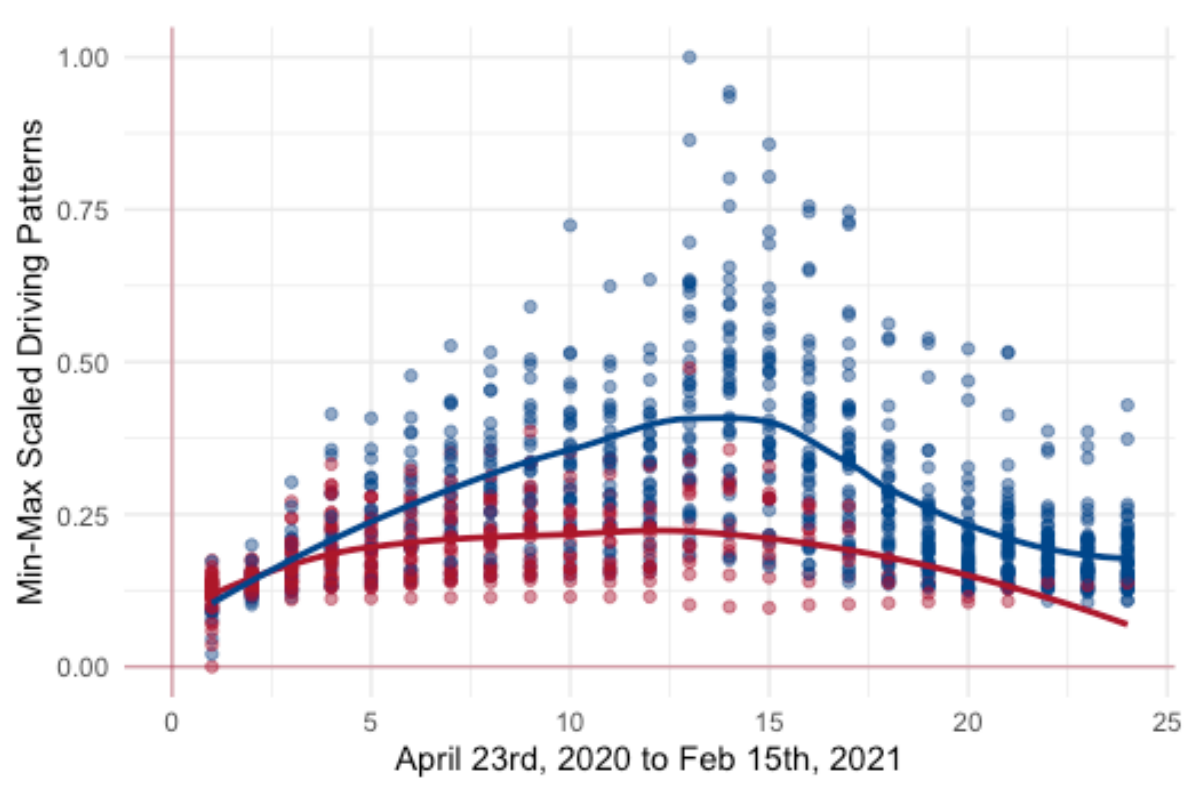

e3B. Google Mobility Retail and Recreation Pattenrs

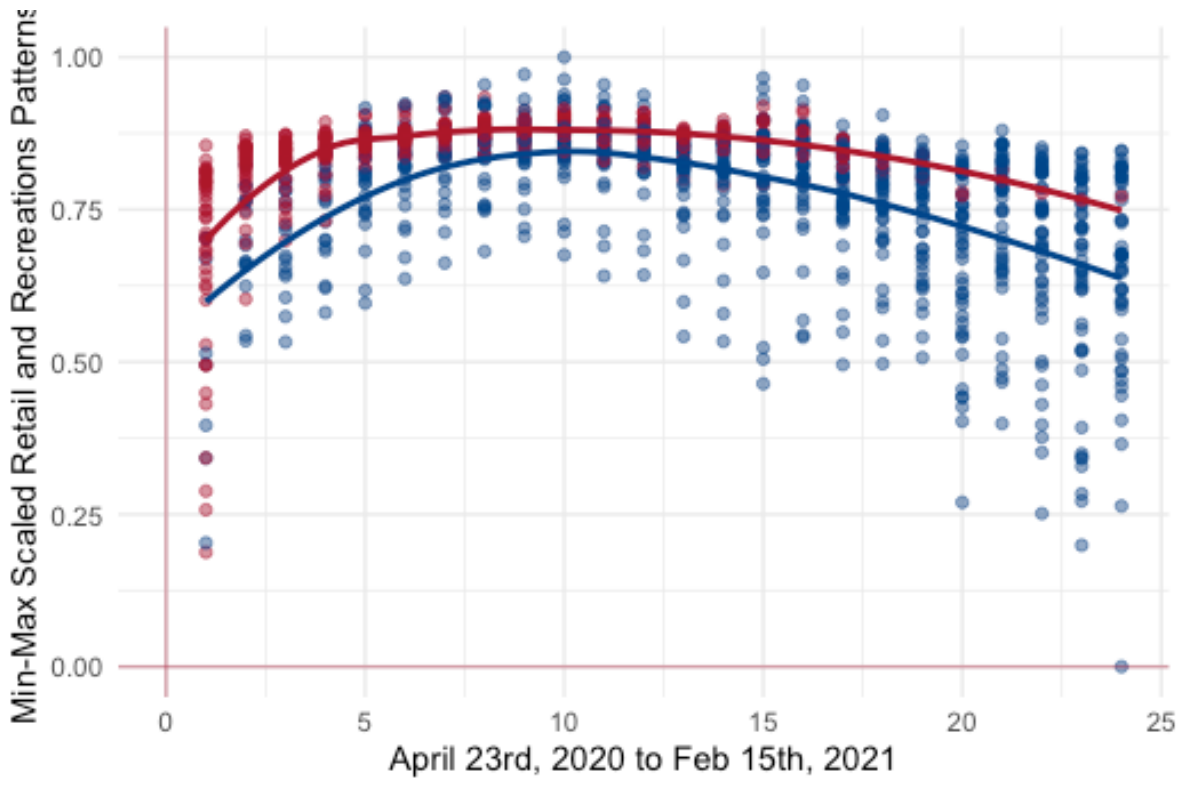

e3C. Google Mobility Grocery and Pharmacy Patterns

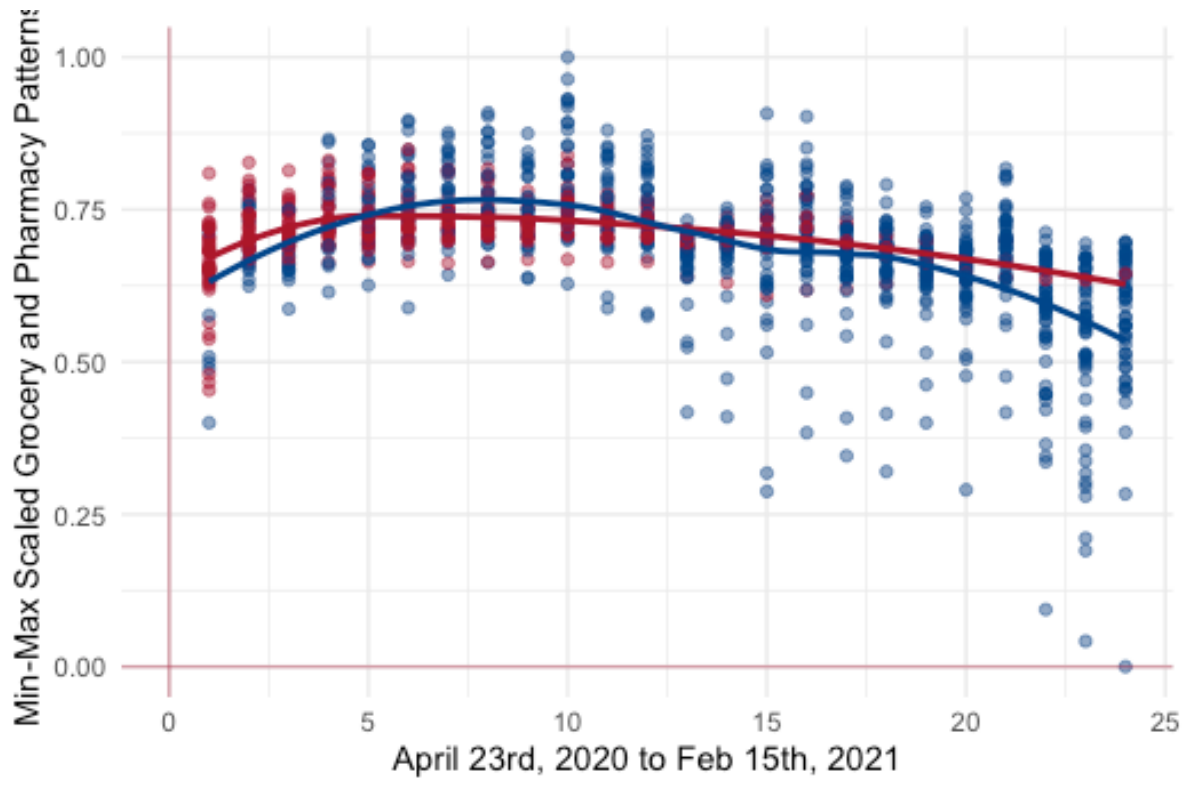

e4D. Proportion Housing Insecure vs. Mental Distress

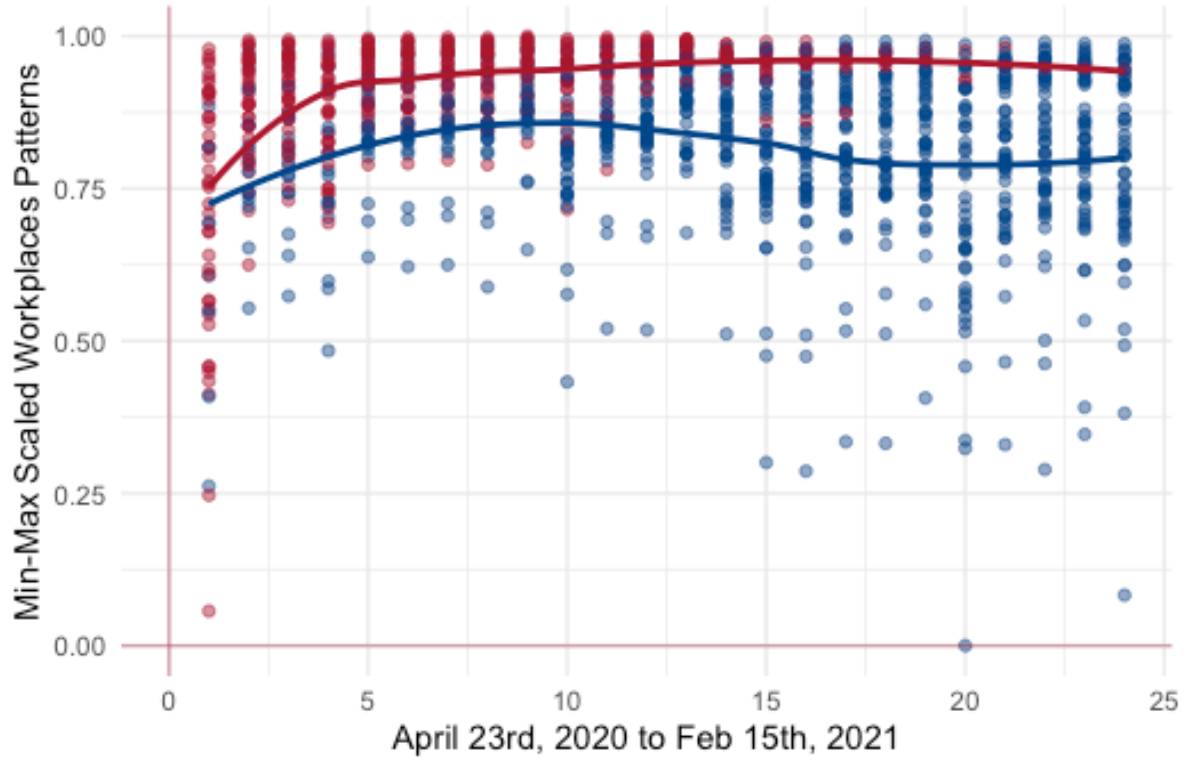

e3D. Google Mobility Residential Patterns

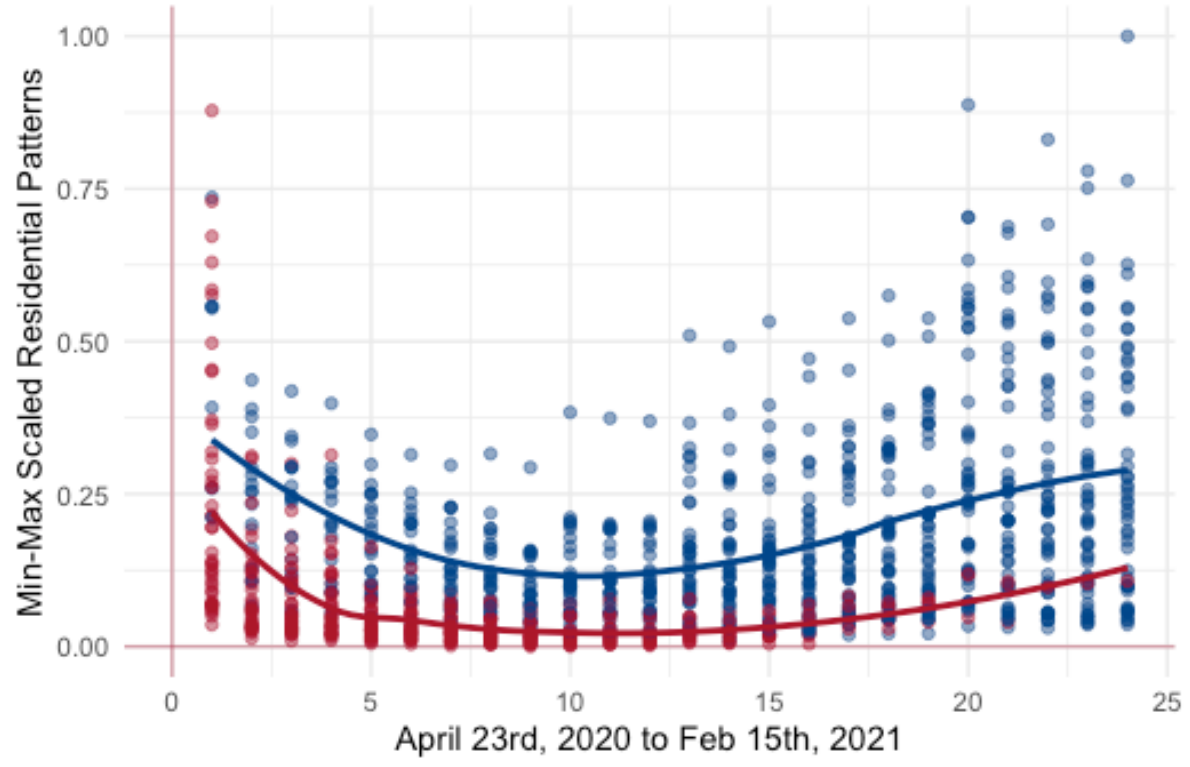

e3E. Google Mobility Parks Patterns

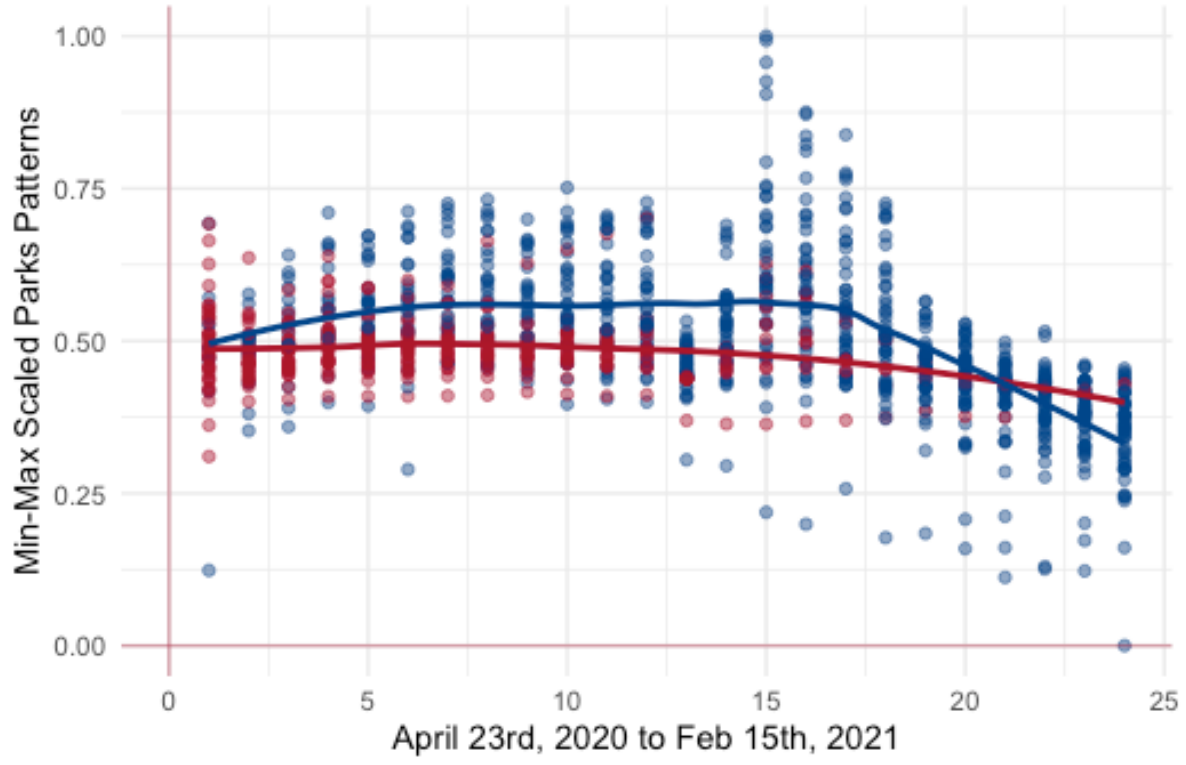

## eTable 7: Correlation between mental health variables, information, and economic distress.

| | $MD_t$ | $GAD2_t$ | $PHQ2_t$ | $Cases_{t-1}$ | $Deaths_{t-1}$ | $Cases_{t-2}$ | $Deaths_{t-2}$ | Housing $Insecure_{t-1}$ | Food $Insecure_{t-1}$ | Job $loss_{t-1}$ |
|---|---|---|---|---|---|---|---|---|---|---|
| $MD_t$ | 1 | | | | | | | | | |
| $GAD2_t$ | 0.971* | 1 | | | | | | | | |
| $PHQ2_t$ | 0.979* | 0.901* | 1 | | | | | | | |
| $Cases_{t-1}$ | 0.429* | 0.417* | 0.418* | 1 | | | | | | |
| $Deaths_{t-1}$ | 0.388* | 0.361* | 0.392* | 0.953* | 1 | | | | | |
| $Cases_{t-2}$ | 0.420* | 0.404* | 0.414* | 0.963* | 0.912* | 1 | | | | |
| $Deaths_{t-2}$ | 0.388* | 0.353* | 0.399* | 0.913* | 0.960* | 0.953* | 1 | | | |
| Housing $Insecure_{t-1}$ | 0.439* | 0.479* | 0.384* | 0.155* | 0.201* | 0.183* | 0.227* | 1 | | |
| Food $Insecure_{t-1}$ | 0.556* | 0.625* | 0.472* | 0.225* | 0.176* | 0.221* | 0.174* | 0.608* | 1 | |
| Job $loss_{t-1}$ | 0.469* | 0.435* | 0.477* | 0.152* | 0.264* | 0.164* | 0.278* | 0.547* | 0.219* | 1 |

Source: Own calculations using HPS data

## eTable 8: Correlation between mental health variables, and behavior

| | $MD_t$ | $GAD2_t$ | $PHQ2_t$ | Transit $Stations_{t-1}$ | $Retail_{t-1}$ | $Grocery_{t-1}$ | $Workplace_{t-1}$ | $Residential_{t-1}$ | $Driving_{t-1}$ | $Parks_{t-1}$ |
|---|---|---|---|---|---|---|---|---|---|---|
| $MD_t$ | 1 | | | | | | | | | |
| $GAD2_t$ | 0.971* | 1 | | | | | | | | |
| $PHQ2_t$ | 0.979* | 0.901* | 1 | | | | | | | |
| Transit $Stations_{t-1}$ | -0.207* | -0.200* | -0.203* | 1 | | | | | | |
| $Retail_{t-1}$ | -0.210* | -0.220* | -0.191* | 0.941* | 1 | | | | | |
| $Grocery_{t-1}$ | -0.183* | -0.200* | -0.160* | 0.729* | 0.783* | 1 | | | | |
| $Workplace_{t-1}$ | -0.151* | -0.160* | -0.136* | 0.732* | 0.816* | 0.588* | 1 | | | |
| $Residential_{t-1}$ | 0.200* | 0.204* | 0.187* | -0.886* | -0.942* | -0.702* | -0.929* | 1 | | |
| $Driving_{t-1}$ | 0.071 | 0.056 | 0.081 | -0.276* | -0.318* | -0.180* | -0.635* | 0.522* | 1 | |
| $Parks_{t-1}$ | -0.070 | -0.126* | -0.019 | 0.210* | 0.218* | 0.547* | -0.021 | -0.088 | 0.385* | 1 |

Source: Own calculations using HPS data

## eTable 9: Correlation between mental health and public health

| | MD | GAD2 | PHQ2 | SAHO | Daycare | Business | Religious | Mask | Restaurant | Job Search Waiver |
|---|---|---|---|---|---|---|---|---|---|---|
| **MD** | 1 | | | | | | | | | |
| **GAD2** | 0.971* | 1 | | | | | | | | |
| **PHQ2** | 0.979* | 0.901* | 1 | | | | | | | |
| **SAHO** | 0.204* | 0.164* | 0.229* | 1 | | | | | | |
| **Daycare** | 0.062 | 0.044 | 0.075 | 0.123* | 1 | | | | | |
| **Business** | 0.353* | 0.294* | 0.387* | 0.192* | 0.051 | 1 | | | | |
| **Religious** | 0.124* | 0.138* | 0.105 | 0.223* | -0.007 | 0.070 | 1 | | | |
| **Mask** | 0.317* | 0.271* | 0.342* | 0.122* | 0.310* | 0.244* | 0.015 | 1 | | |
| **Restaurant** | 0.133* | 0.105 | 0.152* | 0.140* | 0.115* | 0.489* | 0.059 | 0.189* | 1 | |
| **Job Search Waiver** | 0.083 | 0.114* | 0.052 | -0.019 | -0.051 | 0.054 | 0.019 | -0.080 | -0.048 | 1 |

Source: Own calculations using HPS data